\documentclass[5p,twocolumn]{elsarticle}

\usepackage{amsmath}
\usepackage{amssymb}
\usepackage{graphicx}
\usepackage{changes}
\usepackage{epstopdf, epsfig}

\newcommand{\be}{\begin{equation}}
\newcommand{\ee}{\end{equation}}
\newcommand{\bs}{\begin{subequations}}
\newcommand{\es}{\end{subequations}}

\newcommand{\rmd}{\mathrm{d}}
\newcommand{\rmi}{\mathrm{i}}
\newcommand{\rme}{\mathrm{e}}

\newcommand{\half}{{\textstyle\frac1{2}}}

\newcommand{\bU}{\mathbf{U}}

\newcommand{\bk}{\mathbf{k}}
\newcommand{\br}{\mathbf{r}}
\newcommand{\kU}{\bk\cdot\bU}
\newcommand{\bkU}{\kU}
\newcommand{\bV}{\mathbf{V}}
\newcommand{\bee}{\mathbf{e}}

\newcommand{\hu}{\hat{u}}
\newcommand{\hv}{\hat{v}}
\newcommand{\hw}{\hat{w}}
\newcommand{\hp}{\hat{p}}
\newcommand{\hzeta}{\hat{\zeta}}

\newcommand{\Fr}{\mathrm{Fr}}

\newcommand{\pext}{p_\mathrm{ext}}

\newcommand{\kint}{\int\frac{\mathrm{d}^2 k}{(2\pi)^2}}
\newcommand{\zint}{\int\limits_{-h}^{0}\rmd z}

\newcommand{\sqom}{\sqrt{\omega_1^2+\omega_2^2}}

\newcommand{\zetak}{\zeta}
\newcommand{\p}{\partial}
\newcommand{\efac}{\mathrm{e}^{\rmi \mathbf{k}\cdot\mathbf{r}}}

\begin{document}

\begin{frontmatter}

\title{Transient wave resistance upon a real shear current}

\author[ntnu]{Yan Li\corref{cor1}}
\ead{yan.li@ntnu.no}
\author[ntnu]{Benjamin K. Smeltzer\corref{cor}}
\author[ntnu]{Simen {\AA}. Ellingsen}

\cortext[cor1]{Corresponding author.}
\cortext[cor]{YL and BKS are to be considered joint first authors, in alphabetical order.}
\address[ntnu]{Department of Energy and Process Engineering, Norwegian University of Science and Technology, N-7491 Trondheim, Norway}

\begin{abstract}
  We study the waves and wave--making forces acting on ships travelling on currents which vary as a function of depth. Our concern is realism; we consider a real current profile from the Columbia River, and model ships with dimensions and Froude numbers typical of three classes of vessels operating in these waters. To this end we employ the most general theory of waves from free--surface sources on shear current to date, which we derive and present here. Expressions are derived for ship waves which satisfy an arbitrary dispersion relation and are generated by a wave source acting on the free surface, with the source's shape and time-dependence is also being arbitrary. Practical calculation procedures for numerically calculating dispersion on a shear current which may vary arbitrarily with depth both in direction and magnitude, are indicated. 
  
  For ships travelling at oblique angle to a shear-current, the ship wave pattern is asymmetrical, and wave--making radiation forces have a lateral component in addition to the conventional wave resistance, the sternward component. No corresponding lateral force exists in the absence of shear. We consider the dependence of wave resistance and lateral force for upstream, downstream and cross--stream motion on the Columbia River current, both in steady motion and during two different maneouvres: a ship suddenly set in motion, and a ship turning through 360$^\circ$. We find that for smaller ships (tugboats, fishing--boats) the wave resistance can differ drastically from that in quiescent water, and depends strongly on Froude number and direction of motion. For Froude numbers typical of such boats, wave resistance can vary by a factor $3$ between upstream and downstream motion, and the strong Froude number dependence is made more complicated by interference effects. The lateral radiation force is approximately $20\%$ of the wave resistance for cross--current motion for these ships, and can reach more than $50\%$ for short periods during maneouvring; this is by no means a small force, and will have an effect on seakeeping, economy, optimal choice of route and operational safety. For an example ship (tugboat) doing a turning motion, both the lateral force and wave resistance are predicted to undergo variations whose amplitude amounts to approximately $100\%$ of their constant values in quiescent water.   

\end{abstract}

\begin{keyword}
  Wave resistance \sep Shear flow \sep Transient ship waves
\end{keyword}

\end{frontmatter}

\section{Introduction}

Typically, more than 30\% of the fuel consumption of ocean--going ships is from making waves \cite{faltinsen05}. A resistance is felt due to the work done by the ship on the surrounding water, which propagates away in the form of wave energy. While going back over a century \cite{michell1898,havelock09,havelock14,havelock19,havelock22,wehausen73,noblesse83}, wave resistance on ships has also been the focus of recent investigations \cite{benzaquen14}.

Two of us recently showed that the wave resistance acting on a ship in steady motion can be significantly altered by the presence of a shear current beneath the water surface \cite{li16}. In conditions with no shear, wave resistance typically becomes important for Froude numbers around $0.3$ and peaks in the vicinity of $0.5$ before decreasing again as the wake becomes dominated by diverging waves. When a sub-surface shear current is present, however, both the Froude number at which wave resistance sets in, and the value at which it peaks, are in general changed, with opposite effects whether the ship travels along, against, or across the current \cite{li16}. Moreover, sub-surface shear causes the angle made by the ship waves to differ from Lord Kelvin's classic $19.47^\circ$, being smaller for shear-assisted and larger for shear-inhibited motion, and asymmetric around the line of motion when the angle with the current is oblique \cite{ellingsen14a}. 
In the latter case momentum is imparted to the water at different rates to starboard and port, and the corresponding wave radiation force experienced by the ship obtains a lateral component in addition to the conventional sternward wave resistance \cite{li16}. No corresponding phenomenon exists in rectilinear motion if the current has depth-uniform velocity profile.

Our concern in this paper is to introduce realism, compared to previous studies which have considered idealised models. We study how the shear of a real, measured current may affect the wave radiation forces on actual ships. We use an example shear profile measured in the Columbia River delta. These waters are crossed by thousands of ships each year, and we study model ships with dimensions and velocities typical of different vessel types operating there. 
This includes not only the forces acting during steady motion, but also transient forces from manoeuvring motions. To this end, the most general theory of linear ship waves (or waves from free--surface sources more generally) to date has been developed, and is presented here, allowing a shear current to vary arbitrarily with respect to depth both in direction and magnitude, as long as it may be considered uniform in horizontal directions. 

We demonstrate in Section \ref{sec:num} how a real shear current can have a very significant effect on the wave--making forces acting on real ships. At typical Froude numbers we find for smaller boats (tugboats, fishing boats) that the wave resistance can differ by a factor $3$ or more between upstream and downstream motion at the same velocity relative to the free surface. The lateral radiation force acting when travelling across the shear is also very significant; it is typically around $20\%$ of the sternward resistance force in steady motion, but can momentarily reach more than $50\%$ of the wave resistance during maneouvring. These are by no means small effects, and will affect the seakeeping and the optimal choice of velocity and route of travel, and perhaps also cause safety issues for ships manoeuvring in proximity of each other. 

This paper contains two major sections, one theoretical, one of an applied nature. The reader primarily interested in what the practical effect of shear in real--life situations might be, may wish to refer directly to the numerical results in Section \ref{sec:num} bearing in mind the system definitions in Section \ref{sec:def}. The theoretical foundations and framework is laid out in Section \ref{sec:basic}; it has been presented, as far as we have been able to, so as to be useful to readers who wish to employ the formalism for their own purposes.

Studies of transient wave resistance go back a long time. Whenever a ship undergoes changes in velocity during acceleration or manoeuvring, transient waves are emitted, and the wave radiation force correspondingly will be time dependent for the duration during which the created transient ring-wave remains in the immediate vicinity of the ship. A century ago, Havelock studied the wave resistance in 2 dimensions due to a suddenly appearing ship, modelled as a distribution of additional pressure at the water's surface, suppressing the free surface approximately as would a ship \cite{havelock17}. The resistance force was found to increase from zero to a peak value before relaxing in an oscillatory manner to its static value. The speed of relaxation was found to depend closely on the aspect ratio of the disturbance, since the bow and stern waves from a more slender ship tend to cancel, causing a quicker relaxation to steady conditions and a more stable steady wave resistance. On the other hand a circular ``ship'' with little such interference, experienced a very slow relaxation rate. A study of the resistance felt by a submerged cylinder starting suddenly from rest revealed similar results \cite{havelock49}. Studies of ships in various kinds of acceleration is a related classical problem \cite{bhattacharyya56, wehausen60}.

Approaching the problem of waves in three--dimensional systems in the presence of sheared flows, standard methods to calculate waves and motions of floating bodies must be immediately discarded, based as they are on potential theory. No satisfactory theory of creating bodies from submerged sources and sinks exist even in the simplest shear currents exists, not to mention advanced panel methods \cite{ellingsen16}. A feasible approach for our purposes is however to create a ``ship--shaped footprint'' in the free surface by introducing an external surface pressure. The approach goes back over a century \cite{havelock08} and has recently been employed in wave resistance studies \cite{benzaquen14}. Such a model, only affects the dynamic boundary condition, not the equations of motion, thus does not in principle pose any restrictions on the flow vorticity. 

\subsection{Outline}

The investigated system is presented in Section \ref{sec:def} along with the basic formalism. Section \ref{sec:basic} then goes on to develop the general theory of waves from moving, time--dependent surface disturbances upon a horizontal background current which may vary arbitrarily with depth, both in direction and magnitude. In particular, a suitable formalism for working with a general (not explicitly known) dispersion relation is derived in Section \ref{sec:dispersion}, and applied to the general problem in Section \ref{sec:tr}. In Section \ref{sec:approx} practical considerations are presented concerning numerical evaluation of the dispersion relation for arbitrary velocity profiles, and the formalism for calculating wave resistance and lateral radiation force is derived and discussed in Section \ref{sec:R}.

Section \ref{sec:num} is of a more applied nature and presents numerical results for particular situations. A measured velocity profile from the Columbia River estuary is used, and pressure distributions modelling ships of realistic dimensions are employed in order to provide reasonably realistic estimates of the effect of shear in these waters while retaining some generality. For comparison, and to illustrate the effect of shear without the large number of lengthscales and parameters, corresponding results for the simple case of a linearly depth--dependent current are given in Section \ref{sec:numlin} before conclusions are drawn. Some further details on derivation and numerical procedures are found in appendices.

\subsection{System definition}\label{sec:def}

In this section the system under scrutiny is defined, along with general formalism used in the paper. The system is a generalisation of that considered in Ref.~\cite{li16b}.

	\begin{figure}[tb]
	\includegraphics[width=\columnwidth]{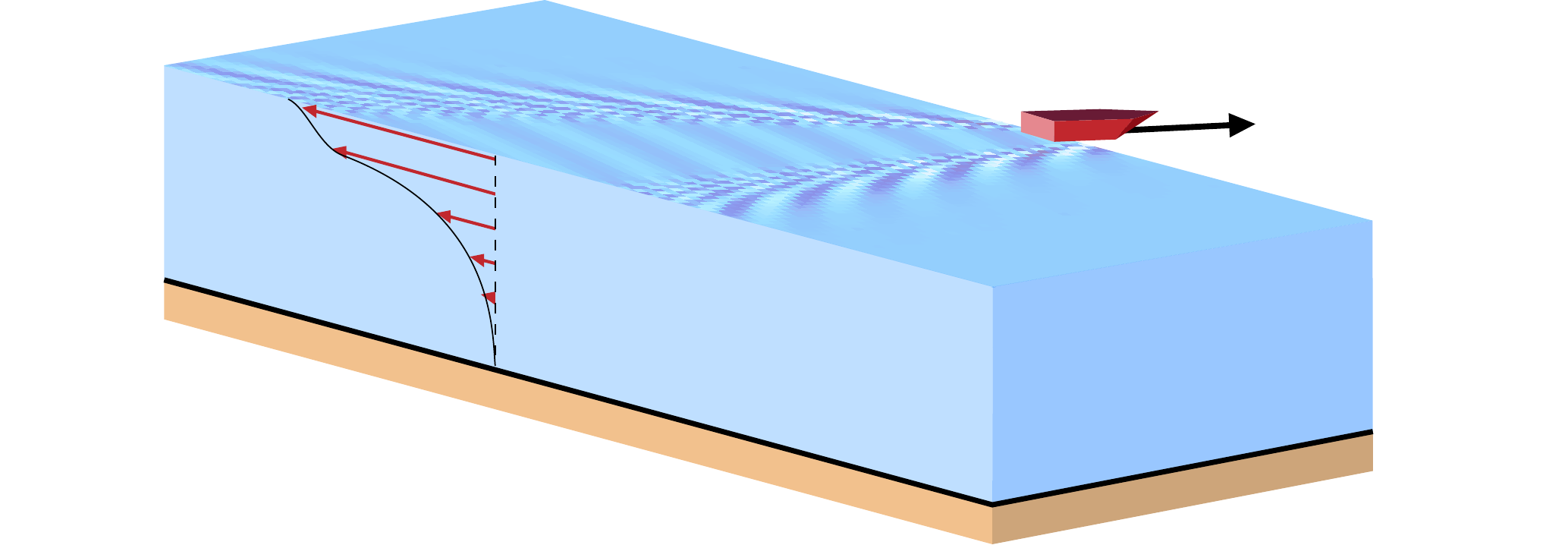}
	\caption{
	  Schematic sketch of the system: a ship travelling with arbitrary, time--dependent velocity atop a shear current of arbitrary depth--dependence. Here a ``lab'' coordinate system is shown, fixed relative to the sea--bed. 
	}
	\label{fig:geom}
\end{figure}

\begin{figure}[tb]
	\begin{center}
		\includegraphics[width=.9\columnwidth]{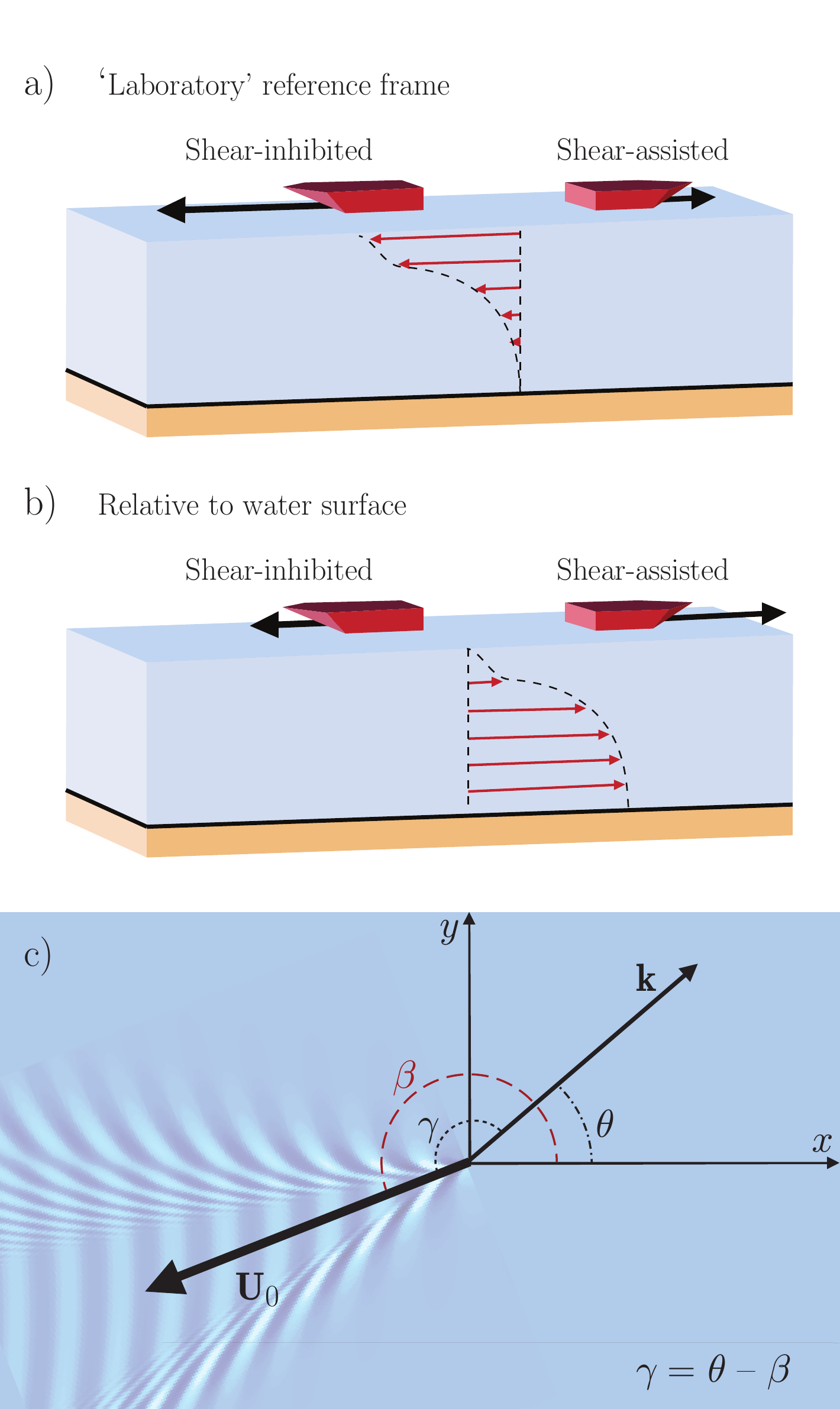} 
	\end{center}
	\caption{(a) Illustration of shear--assisted vs shear--inhibited ship motion; shown in the ``lab'' reference frame relative to the sea bed, and (b) in the reference frame relative to the water surface. (c) Definition of angles 
	$\gamma$ (angle between $\bk$ and $\bU_0$), $\beta$ (angle between $\bU_0$ and $x$ axis, and $\theta$ (angle between $\bk$ and $x$-axis). The reference frame is here at rest with respect to the ship. Note: $\beta=0$ is the maximally shear assisted direction of motion, $\beta=\pi$ the maximally shear inhibited.}
	\label{fig:agls}
\end{figure}

We consider infinitesimal wave amplitudes described by the surface elevation function $ \hzeta(\br,t) $ with horizontal position $\br=(x,y)=r(\cos\varphi,\sin\varphi)$ and time $t$.    
The waves are superimposed on a depth-varying background flow $\mathbf{U}(z)$. In our general theory in Section \ref{sec:basic}, $\bU(z)$ may vary both in magnitude and direction, although our numerical examples in Section \ref{sec:num} will all be unidirectional. We use the shorthand $\bU(0)=\bU_0$. A sketch of the system is seen in Fig.\ref{fig:geom}. We assume incompressible and inviscid flow. 
The three velocity components and pressure perturbation due to the waves we name  $\hu,\hv,\hw$, and $\hp$, respectively, all functions of $\br,z$ and $t$. Hatted quantities are considered small, and we linearise with respect to these. The flow field is thus $[\bV,P]=[\bU(z)+\hu\bee_x+\hv\bee_y+\hw\bee_z,-\rho gz+\hp]$, with $\bV$ and $P$ the total velocity and pressure fields, respectively, $g$ the gravitational acceleration, and $\rho$ the density of the water.
The flow obeys the Euler equation
\be\label{euler}
  \partial_t\bV + (\bV\cdot\nabla)\bV=-\nabla P/\rho - g\mathbf{e}_z.
\ee
We neglect surface tension. 
The physical quantities are defined in Fourier space of the horizontal plane as $[\hzeta,\hu,\hv,\hw,\hp](\br,z,t)\leftrightarrow [\zeta,u,v,w,p](\bk,z,t)$ as
\be\label{fourier}
  [\hzeta,\hu,\hv,\hw,\hp](\br,z,t)=\kint [\zeta,u,v,w,p](\bk,z,t)\efac
\ee
so that  $ \mathbf{k}=(k_x,k_y)=(k\cos\theta,k\sin\theta) $ is the wave vector (It is understood that $\hzeta$ and $\zeta$ do not depend on $z$).
The water depth $h$ is constant, and may be allowed to tend to $\infty$. 

In the system sketched in Fig.~\ref{fig:geom} no less than three different reference frames are natural, depending on the question under consideration. Fig.~\ref{fig:geom} shows the ``lab'' reference frame, i.e., as seen by an observer on shore. A second frame of reference which we use in Section \ref{sec:tr} is that which is fixed on the moving model ship. Finally, in section \ref{sec:num} we will sometimes work in the frame of reference in which the water surface is at rest. 

For this reason the oft used terms `upstream' and `downstream' are ambiguous as denotations of directions of motion. 
We will instead use the terms `shear--assisted' and `shear--inhibited' to describe directions of ship motion or wave motion relative to the sub--surface current. The motion is assisted by the current if, in a reference system where the water surface is at rest, the sub-surface current has a component along the direction of motion (this corresponds to the ship travelling upstream in the case of e.g.\ a river). Correspondingly, for shear--inhibited motion the sub--surface current has positive component against the ship's motion, in a system where the free surface is at rest (corresponds to downstream motion on a river). These concepts are visualised in Fig.~\ref{fig:agls}a and b. They are only strictly well defined only for velocity profiles that do not change direction or sign relative to the free surface, yet this is sufficient for our present purposes.

In later sections we shall make use of polar coordinates in the horizontal plane, which we define in figure \ref{fig:agls}c, for a system in which the ship is at rest. Note that the angle $\beta$ differs by $\pi$ from that used in \cite{ellingsen14a,li16}, where a reference system relative to the water surface was used. 
The angle between $\bk$ and $\bU_0$ is $\gamma$.

  \section{Theory: linear surface waves from an arbitrary time--varying wave source, propagating on an arbitrary shear current}
\label{sec:basic}

In this section we present a theoretical framework for calculating waves from arbitrary wave sources on the free surface, in flows with arbitrary dispersion relation $\omega(\bk)$, affected by sub-surface currents that may vary both with depth and direction. To our knowledge no theory this general has ever been presented. As a special case the theory provides a procedure for calculation and analysis of ship waves on arbitrary horizontal shear currents. 

From the linearised Euler equations and continuity equation in $\bk$-space we have the relations (cf.\ e.g.\ the procedure of \cite{shrira93})
\begin{subequations}
\begin{align}
	(\partial_t+\rmi \kU)w'(z,t)-\rmi \kU'w(z,t) =& -k^2 p(z,t)/\rho, \label{eq:p}\\
	(\partial_t+\rmi \kU)w(z,t) =& - p'(z,t)/\rho, \label{eq:dp}
\end{align}
\end{subequations}
where a prime denotes differentiation with respect to $z$, and the dependence on $\bk$ of $p$ and $w$ is suppressed here and henceforth.

\subsection{General form of surface wave dispersion relation}
\label{sec:dispersion}

We will now present a general, implicit form of the dispersion relation for waves atop a general depth--dependent shear flow $\bU(z)$. The relation allows us to derive general expressions for surface waves from an arbitrary free--surface source in Section \ref{sec:tr}. Determining $\omega(\bk)$ for a specific situation is the topic of Section \ref{sec:approx}.

We use the physical values $\omega_\pm(\mathbf{k})$ to express the free--surface elevation for a given $\mathbf{k}$-component as:
\be\label{zeta}
  \zetak(\mathbf{k},t) = Z_+(\mathbf{k})\mathrm{e}^{-\rmi \omega_+ t} + Z_-(\mathbf{k})\mathrm{e}^{-\rmi \omega_- t}
\ee
where $Z_\pm$ are unknown coefficients to be determined. 
Also the other perturbed quantities $u,v,w$ and $p$ will have time dependence $\propto \exp(-\rmi \omega_\pm t)$. 

If the values of $Z_\pm$ are known from initial conditions, the full time dependent solution to the free--surface elevation can be found from \eqref{zeta}. 

The phase velocities $\omega_+(\bk)/k$ and $\omega_-(\bk)/k$ correspond to partial waves propagating in directions $\bk$ and $-\bk$, respectively. They satisfy the relation
\be\label{omrel}
  -\omega_-(\bk)=\omega_+(-\bk).
\ee
Hence there is a unique, positive phase velocity $\omega_+(\bk)$ in propagation direction $\bk$, and the integral over all $\bk$ effectively accounts for each mode twice. The relation \eqref{omrel} is general and holds for any shear current. 
We show in \ref{appx-1} that the dispersion relation for a plane wave of small amplitude on a depth--dependent flow may be written
	\begin{align} \label{dispR}
	       	\Delta_R&(\bk, \omega) \equiv   (1+I_g) (\omega - \kU_0)^2 + \notag\\
	       	& (\omega - \kU_0)\kU'_0{\tanh kh}/{k}-gk \tanh kh = 0,
	\end{align}
where $\Delta_R$ is defined for later reference, and
\be
  I_g(\bk)  =\zint \dfrac{\kU''(z) w(z,0) \sinh k(z+h)}{k[\kU(z)-\omega] w(0,0)\cosh kh}. 
\ee
The implicit dispersion relation \eqref{dispR} is extremely useful for analytical purposes. It is not itself closed, since both $\omega(\bk)$ and $w(z,t)$ are unknowns. 
The two roots of the equation $\Delta_R=0$ are $\omega=\omega_\pm(\bk)$. It is found e.g.\ in \cite{smeltzer17} that the zeros of $\Delta_R$ are simple, hence Eq.~\eqref{dispR} may be written on the form
\be \label{Domg}
	\Delta_R(\bk, \omega) = (1+I_g)  (\omega-\omega_+)(\omega-\omega_-)=0. 
\ee

\subsection{Waves from an arbitrary, time-dependent pressure distribution} \label{sec:tr}

We wish to find a solution to the surface pattern 
resulting from a time-dependent externally applied pressure distribution $\hat{p}_\mathrm{ext}(\br,t)\leftrightarrow\pext(\bk,t)$ at the free surface. 

The pressure, when positive, depresses the water surface thus modelling a moving wave source such as a ship. Using an applied surface pressure as wave source rather than e.g.\ potential theory with submerged sources such as are often used in the theory of ship motions \cite{faltinsen90}, is advantageous since only the boundary conditions are directly affected. This is necessary in our system, since the flow we consider is inescapably rotational and potential theory is inapplicable. 
It should be noted that the relation between the shape of the applied pressure and the resulting surface depression is not altogether trivial for a moving source, and has some Froude number dependence. This introduces a certain quantitative uncertainty in the results presented in section \ref{sec:num}; this is a question we intend to address in the near future. 

By superposition, the response $G(\bk,t)$ of the system to an arbitrary time-dependent pressure distribution can be expressed as a time-integral of pressure pulses emitted at all previous times,
\be\label{evolution}
  G(\mathbf{k},t) = \int_{-\infty}^t \rmd \tau \pext(\mathbf{k},\tau)H(\mathbf{k},t-\tau).
\ee
$H(\bk,t-t_0)$ is the system's response to an impulsive pressure rate 
$p_I(t) = I\delta(t)$ 
which imparts a finite impulse to the free surface during an infinitesimally short time.
$I$ equals unity in units of pressure. $G$ and $H$ physically may represent any of the perturbation quantities $u,v,w,p$ or $\zeta$. 
Mathematically $H$ plays the role of a Green's function.

We now proceed to finding the response of the free surface to a pressure impulse. In Eq.~\eqref{evolution} we let $G\to\zeta$, and the correspondng response function we call $H_\zeta(\bk,t)$. 
The full time evolution $\zeta(\bk,t)$ for $t>0$ is then calculated from \eqref{evolution} as
\be\label{zetat}
  \zeta(\bk,t)=\int_{-\infty}^t \rmd \tau \pext(\bk,\tau)H_\zeta(\bk,t-\tau)
\ee
with $H_\zeta$ derived in the following, given in \eqref{Hz}.

The prescribed impulsive pressure enters the equation system via the dynamic free surface boundary condition, which can be written
\begin{align}
  \rmi \kU^\prime_0w - \left(\p_t + \rmi \kU_0 \right) w^\prime - k^2g\zeta & = k^2I\delta(t)/\rho.\\
\left(\p_t + \rmi \kU_0 \right) \zeta &  =  w, \label{kbc}
\end{align}
with $w,w'$ evaluated at $z=0$. 
Here $\bU_0$ is surface velocity, and a prime denotes differentiation with respect to $z$.
Integration over an infinitesimal time interval 
$t = 0_-$ to $0_+$ yields the following relations for $w(z,t)$ and $\zeta(t)$,
\bs \label{eq:t=0}
\begin{align}
  w^\prime(0,0_+) & = -k^2I/\rho, \label{wI}\\
  \zeta(0_+)                  & = 0, \\
  \dot{\zeta}(0_+)        &= w(0,0_+) ,
\end{align}
\es
using the assumptions that the system is completely at rest for 
$t<0$ and that all physical quantities have finite values at $t>0$, at $ t=0_+ $ in particular.
We suppress the dependence of $w$ and $\omega$ on $\bk$ in this subsection.

When a current of arbitrary depth--variation is present, the primary challenge is that analytical expressions for $\omega_\pm(\bk)$ and $ w(z,t) $ cannot be found. We show in \ref{appx-1} the relations
\bs\label{eq:dwF}
\begin{align} 
     w^\prime(0,0_+) & = k(1+I_g )w(0,0_+)\coth kh, \\
                    & = -\dfrac{\kU^\prime_0\tilde{\omega}+gk^2}{\tilde{\omega}^2}w(0,0_+),\\
                    &= \frac{k}{F(\bk)}w(0,0_+), \label{Fdef}
\end{align}
\es
where $ \omega $ can be either of the roots of $\Delta_R=0$, i.e.\ $ \omega_+ $ or $ \omega_- $, and the intrinsic frequency is $\tilde{\omega}=\omega - \kU_0$. 
Eq.~\eqref{Fdef} defines the quantity $F(\bk)$ for later reference. 
We note that $F(\bk)$ can be written in several different forms,
\bs
\begin{align}\label{Falt}
  F(\bk) &= \frac{k w(0,0_+)}{w'(0,0_+)}\\
  &=\frac{\tanh kh}{1+I_g}\label{Fth}\\
  &=\frac{(\omega-\omega_-)(\omega-\omega_+)}{\Delta_R}\tanh kh\\ 
  &=\frac{k\tilde{\omega}(\bk)^2}{gk^2-\kU_0'\tilde{\omega}(\bk)}.\label{Fom}
\end{align}
\es
Which form of $F(\bk)$ is most convenient is different in different cases. The final form \eqref{Fom} has the advantage that only the value of $\omega(\bk)$ is required when $\bU(z)$ is known.

From \eqref{zeta}, \eqref{eq:t=0} and \eqref{eq:dwF} then follows
\begin{subequations}
\begin{align}
  Z_+ + Z_- =&0;\\
  \omega_+Z_+ + \omega_-Z_- =& -\rmi Ik  F(\bk)/\rho
\end{align}
\end{subequations}
Solving for $Z_\pm$ and inserting into 
\eqref{zeta} yields the surface elevation $H_\zeta$ from an impulsive pressure pulse as
\be \label{Hz}
  H_\zeta(\mathbf{k},t) = \frac{\rmi k  F(\bk)
  }{2\rho\omega_\text{div}(\bk)}(\rme^{-\rmi\omega_-t}-\rme^{-\rmi\omega_+t}),
\ee
where  the ``divergence frequency'' is, using \eqref{omrel},
\be
  \omega_\text{div}(\bk) = \frac12[\omega_+(\bk) - \omega_-(\bk)]=\frac12[\omega_+(\bk) + \omega_+(-\bk)],
\ee
so that $\omega_\text{div}/k$ is the phase speed with which oppositely propagating waves move apart. 
%

\subsubsection{Suddenly appearing ship}\label{sec:sudden}

As a step towards modelling a ship during manoeuvring  or acceleration in a simple manner, we
consider the special case where $\pext$ is constant for $t>0$ and zero at $t<0$, i.e., a ``ship'' that is launched at $t=0$ already having its final velocity and continuing in steady motion thereafter. This is the system considered long ago by Havelock \cite{havelock17}. It is an artificial situation, but one which can be used as a building block to model more realistic situations. Turning the arrow of time yields instead a suddenly disappearing ship, and adding at the same instance the appearence of the same ship but with a slightly different velocity, say, is a simple model of a rapidly turning and/or accelerating ship. In numerical examples we will consider the more realistic case of a suddenly starting ship.

We use a reference frame following the ship, so that the motion of the ship relative to the water surface is contained in the surface current velocity $\bU_0$ as measured in this system.
The time integral in \eqref{zetat} can be solved explicitly, and $\zeta$ splits naturally into a steady and a transient contribution
\bs\label{generalz}
\begin{align}
  \hzeta(\br,t) =& \lim_{\epsilon\to 0}[\zeta_s(\br) + \zeta_t(\br,t)],\\
  \hzeta_s(\br) =&\frac1\rho\kint \frac{k\pext(\bk)  F(\bk)  }{(\omega_+-\rmi\epsilon)(\omega_--\rmi\epsilon)}\efac,\label{staticGen}\\
  \hzeta_t(\br,t)=&\frac1\rho\kint \frac{k\pext(\bk)  F(\bk)  \efac}{2\omega_\text{div}(\bk)}\notag \\
  &\times\left(\frac{\rme^{-\rmi \omega_+ t}}{\omega_+-\rmi\epsilon}\right.\left.-\frac{\rme^{-\rmi \omega_- t}}{\omega_--\rmi\epsilon}\right). \label{eq:tranW}
\end{align}
\es
Subscripts $s$ and $t$ denote stationary and transient, respectively. Upon splitting into $\zeta_s$ and $\zeta_t$ it was necessary to employ a radiation condition by adding a small imaginary part $-\rmi\epsilon$ to wave frequencies, whereby $\omega_\pm\to\omega_\pm-\rmi \epsilon$ (see, e.g., \cite{li16})  
assuring that waves can only be radiated away from the source. Mathematically this moves the poles to complex values of $\bk$, rendering the integrals definite. Physically, it introduces an arrow of time by implying the time--independent $\hzeta_s$ was ``switched on'' some time in the far past, and consequently likewise the transient contribution which exactly cancels the steady one for $t<0$.

Given a value for $\omega_\pm(\bk)$ (using any of various approximation schemes described below), equation \eqref{zetat} now produces $\hzeta(\br,t)$ at all times; the Fourier transform is taken as in equation \eqref{fourier}, for example using a fast Fourier transform (FFT) algorithm.

\subsubsection{Stationary ship waves}
The simplest case is the classical situation of a ship which has been travelling at constant velocity for a long time. 
The wave pattern in this case is readily obtained from \eqref{zetat} when taking the limit $ t \to \infty $, which yields
\begin{align}
	\hzeta(\br)&=\hzeta_s(\br)\notag \\
	&= \lim_{\epsilon\to0}\kint\frac{k\pext(\bk) \tanh kh }{\rho\Delta_R(\bk,\omega+\rmi \epsilon)}\efac,\label{shipW}
\end{align}
Using $\Delta_R$ on the form \eqref{Domg} is instructive. 
Transient waves described by \eqref{eq:tranW} vanish at large times $ t\to\infty $, as will be further discussed in \S\ref{sec:asymp}. 
Eq.~\eqref{shipW} is exactly the expression for ship waves from a ship moving with velocity $-\bU_0$ relative to the water surface, as derived in \cite{li16} (note that angle $\beta$ differs by $\pi$ from that of \cite{li16,ellingsen14a}), generalised to the case of general dispersion.

\subsubsection{Suddenly starting ship: wave patterns and asymptotics}
\label{sec:asymp}

\begin{figure}[thb]
  \includegraphics[width=\columnwidth]{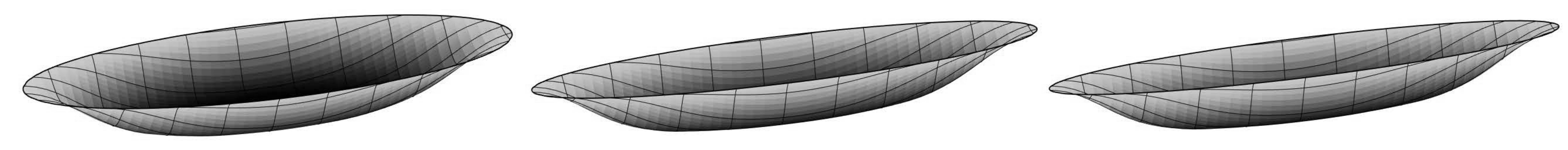}
  \caption{
  Super-Gaussian model ship pressure distributions from Eq.~(\ref{p}). Aspect ratios (left to right) $W=3,5,8$.
  }
  \label{fig:hulls}
\end{figure}

\begin{figure*}[htb]
  \begin{center}
  \includegraphics[width=.95\textwidth]{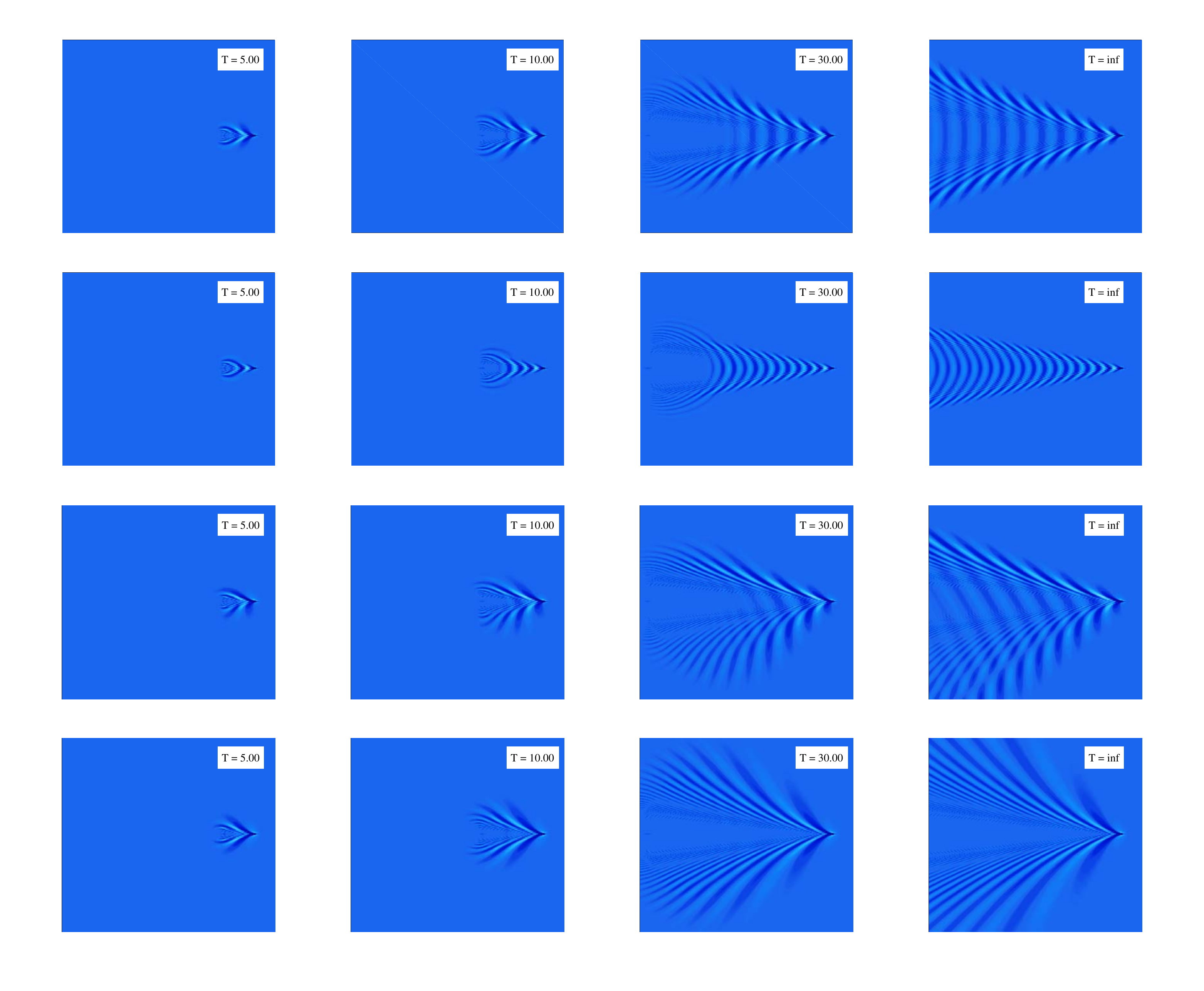}
  \end{center}
  \caption{
    Wave patterns of model ship suddenly set in motion from rest at $T=0$, at increasing nondimensional time $T=t\sqrt{g/L}$ where $L$ is the ship length. The ship is modelled as a super--Gaussian of aspect ratio $L/b=6$; see Eq.~\eqref{p}.   First row: no shear;    Second row: shear--assisted ($\beta = 0$); Third row: side--on shear ($\beta=\pi/2$); Fourth row: shear--inhibited ($\beta=\pi$).   The shear Froude number is $\Fr _s=S|\bU_0|/g=0.8$ with $\bU_0$ the ship velocity relative to the water surface. The reference system is relative to the ship, rotated so that ship motion is the same in all cases.
  }
  \label{fig:patterns}
\end{figure*}

We consider now the model of a ship which starts suddenly from rest. Formally this situation is created from the ``suddenly appearing ship" model in Section \ref{sec:sudden} by superposing the ring wave from a ship at rest suddenly disappearing at $t=0$, and reappearing in the same instance with velocity $\bU_0$ relative to the water surface. 

As a simple model ``ship'' we use an elliptical super-Gaussian pressure distribution with length $L$ and beam (width) $b$ of the form 
\be\label{p}
  \pext(\br,t) = p_0\exp\left\{ -\pi^2\left[\left(2x_\beta/L\right)^2 + \left(2y_\beta/b\right)^2\right]^3 \right\},
\ee
where
\bs
\begin{align}
  x_\beta(t) &= [x-x_0(t)]\cos\beta(t) + [y-y_0(t)]\sin\beta(t), \\
  y_\beta(t) &= -[x-x_0(t)]\sin\beta(t) + [y-y_0(t)]\cos\beta(t)
\end{align}
\es
expressed along the major and minor axes of the ellipse in a reference system (e.g.\ relative to the water surface) where the ship's position may be time--dependent.
In a reference system fixed on the ship, $x_0=y_0=0$ and $[x_\beta,y_\beta] = r[\cos(\varphi-\beta),\sin(\varphi-\beta)]$. The Froude number is $\mathrm{Fr} = |\bU_0|/\sqrt{gL}$.
The super-Gaussian is a fairly realistic model of the submerged part of a hull shape, while avoiding having to specialise to a particular type of hull. Model ``ship'' pressure distributions for some aspect ratios are shown in Fig.~\ref{fig:hulls}.

When first set in motion, the ship creates an initial ring wave which propagates away. After some time the transient ring wave, $\zeta_t$, has disappeared from sight and only a stationary ship wave pattern behind the travelling ship, $\zeta_s$, remains. This is clear from Fig.~\ref{fig:patterns}, where the wave patterns are shown for increasing times after appearence, for different directions of motion atop a linear shear profile 
in deep water.

For large times the transient surface wave $\hzeta_t$ at some point far from the origin will vanish as $t^{-1/2}$. This can be shown rigorously with path integral methods and the stationary phase approximation, but is also physically clear from noting that the full transient wave energy will eventually radiate through any vertical, circular control surface of radius $R$, and wave energy must thus fall off as $R_r(r)^{-1}$ for a ring wave of radius $\sim R_r$. Since wave energy of each Fourier mode moves outward in the far-field at a constant, $k$-dependent group velocity, $R_r\sim c_g t$, and since wave energy is $\propto \hzeta^2$, the time dependence $\hzeta_t\sim t^{-1/2}$ follows for large $t$.

\subsection{Practical calculation techniques for arbitrary velocity profiles}
\label{sec:approx}

To calculate the free--surface elevation 
\eqref{zetat} 
one needs to find the roots $\omega_\pm(\bk)$ of \eqref{dispR}, which is itself not closed since both $ \omega $ and $ w(z) $ are unknowns. Analytical results are in general not available, except for the simplest current varying linearly with depth. 
There are several numerical or semi--analytical techniques that allow calculation of $\omega_\pm(\bk)$ for an arbitrary $\bU(z)$ which we briefly review in this section.

\subsubsection{Simplest case: linear profile}\label{sec:linear}

Consider first the simplest case of a linearly depth--dependent current. This is the only known case where an explicit, analytical dispersion relation is available for all $\bk$. This idealised case is therefore instructive for analysis since analytical results can be derived. 


To calculate ship waves during steady motion, say, one might work in a frame of reference where the model ship is at rest, and the ship's velocity relative to the water surface is $-\bU_0$ where $\bU_0=[U_0,V_0]=|\bU_0|[\cos\beta,\sin\beta]$ (see also Fig.~\ref{fig:agls}c). The current is unidirectional, i.e., 
$\bU(z)=\bU_0+S z\mathbf{e}_x$. 
(This corresponds a ship moving in direction $\beta+\pi$ relative to the water surface. )
We define \cite{ellingsen14a}
\be
  \Fr _s=\frac{|\mathbf{U}_0|S}{g}.
\ee

For the linear shear profile one obtains \cite{li15a,ellingsen14b}
\begin{subequations}\label{linrel}
\begin{align}
  \omega_\pm =& \omega_1 \pm \sqom;\\
  \omega_1 =& \kU_0 -\half S\tanh kh\cos\theta;\\
  \omega_2^2 =& (S\cos\theta\kU_0 +gk)\tanh kh -(\kU_0)^2,
\end{align}
\end{subequations}
hence $\omega_+\omega_- = -\omega_2^2$, and
\begin{align}\label{linrel2}
  \omega_\text{div} &= 
  \sqrt{gk\tanh kh+(S \tanh kh\cos\theta/2)^2}.
\end{align}
Since $I_g=0$ when $\bU''(z)=0$, \eqref{Fth} simply gives $F(\bk)=\tanh kh$. Determining $F$ and $\omega_\pm$ is sufficient for calculating all cases considered above, the most general case being \eqref{zetat} with \eqref{Hz}.

\subsubsection{The piecewise--linear approximation} \label{sec:pla}

A useful numerical scheme to this end is the piecewise--linear approximation (PLA), which was analysed in Refs.~\cite{smeltzer17,zhang05}, and which we will use herein to obtain numerical results. As described herein the PLA is restricted to unidirectional $\bU(z)$; extension to shear currents changing direction is relatively straightforward. Alternative approximations to the dispersion relation are thereafter briefly discussed in section \ref{sec:altapprox}. 

The piecewise--linear approximation (PLA), sometimes called the $ N-$layer model, utilises the fact that explicit solutions are available when the velocity profile is linear as discussed above. A smooth velocity profile $u(z)$ is approximated by a series of linear segments inside $N$ artificial layers, allowing the solution to the vertical velocity to be expressed explicitly within each layer and matched at the artificial layer boundaries. We provide further details in \ref{appx-2}. Following the derivation process in \cite{smeltzer17}, within the top layer the vertical velocity satisfies 
\begin{align} \label{eq:w1}
  w (\bk,t) 
    = & A_1 \sinh k(z+h_1) + B_1 \cosh k(z+h_1), \notag \\
                       & \text{for} -h_1<z<0,	
\end{align}
in which $ h_1 $ is the thickness of the top layer and $ A_1$ and $ B_1 $ are coefficients depending on $\bk$ and $t$, which are determined by the matching conditions at the $N-1$ layer interfaces and from free--surface and bottom boundary conditions. Inserting \eqref{eq:w1} into the first form of $ F(\bk) $ in \eqref{Falt} yields 
\be\label{FPLA}
F(\bk) = \dfrac{A_1\sinh k h_1 + B_1\cosh k h_1}{A_1\cosh k h_1 + B_1\sinh k h_1}
\ee
evaluated at $t=0$.

The next essential step is to obtain solutions for $ \omega_\pm $, exact or approximate, and to determine $ A_1 $ and $ B_1 $ via the PLA procedure \cite{smeltzer17}. The PLA is particularly suitable for problems which are solved in the Fourier plane since it provides a rapid and accurate solution to the dispersion relation $\omega(\bk)$ equally well for all wavelengths, converging to the exact value as $N$ increases \cite{smeltzer17,zhang05}. For our numerical demonstrations we find that $4$-$5$ layers are typically enough at the $~1\%$ accuracy level. 

\subsubsection{Alternative approximations to the dispersion relation}\label{sec:altapprox}

A simpler approach than the PLA can be obtained by evaluating $\omega_+(\bk)$ using an explicit, approximate dispersion relation. The accuracy of such approximations is not so easily predicted, however, and is different in different areas of the $\bk$ plane. 
A much used approximation which is accurate to within a few percent for all $\bk$ in many cases, is the relation by Kirby \& Chen \cite{kirby89}
\be\label{kirby}
  \omega_+(\bk) \approx \omega_0(k) + \int_{-h}^0\rmd z\frac{2\bk\cdot\bU(z)\cosh 2k(z+h)}{\sinh 2kh}
\ee
where $h$ is the total depth of the flow and $\omega_0=\sqrt{gk\tanh kh}$ (note that this 3D generalization of the Kirby \& Chen expression also allows the direction of $\bU$ to vary with $z$). 
The approximate value for $\omega_+(\bk)$ is inserted into equations \eqref{zetat} via \eqref{Fom}. 

We recently made progress on the question of analytical approximations to dispersion relations, deriving error estimates for \eqref{kirby} and also presenting a more robust alternative to \eqref{kirby} in Ref.~\cite{ellingsen17}. Two of us (YL \& S\AA E) have also developed and implemented another numerical method, a simple and promising alternative to the PLA based on direct integration of \eqref{eq:p} and \eqref{dispR} (manuscript in preparation).

\subsection{Transient wave resistance and radiation force}\label{sec:R}

A travelling ship imparts momentum to the water around it to create waves, giving rise to a wave radiation force acting on the ship in the opposite direction. In the absence of shear the wave radiation force always points sternwards for ships in rectilinear motion, and is called wave resistance, or wave--making resistance. Wave resistance typically accounts for more than 30\% of the energy consumption of ocean going vessels \cite{faltinsen05}.

We work in a reference frame where the ship is at rest, and the water surface moves at velocity $\bU_0$ as shown in Fig.~\ref{fig:agls}c. 
Following Havelock \cite{havelock17} the wave radiation force created by a travelling pressure distribution is the force exerted by the external pressure $\hp_\text{ext}(\br,t)$ acting on vertical projections of the moving surface $\hzeta(\br,t)$. The force along unit vector $\bee_f$ acting on horizontal area $\rmd^2 r$ at $\br$ is thus
\be
\rmd f(\br,t) = \hp_\text{ext}(\br,t)(\bee_f\cdot\nabla)\hzeta(\br,t)\rmd^2 r.
\ee 
A ship travelling at an oblique angle with a sub--surface shear current will in general radiate waves asymmetrically around its line of motion, and the radiation force will consequently have both a sternward and a lateral component. The two components are derived with the methods laid out in \cite{li16}, to yield
\begin{align}\label{R}
  \begin{array}{c}R_{\parallel}(t)\\R_{\perp}(t)\end{array}=&-\frac 1{U_0} \int \rmd^2r \hat{p}_\text{ext}(\br,t)
  \left(\begin{array}{c}\bU_0\\\mathbf{e}_z\times\bU_0\end{array}\right)\cdot\nabla \hzeta(\br,t)\notag\\
  =&-\rmi\kint 
 \left(  \begin{array}{c}k\cos\gamma\\k\sin\gamma\end{array}\right) 
  \pext^*(\bk,t)\zeta(\bk)
\end{align}
where an asterisk denotes the complex conjugate and $\parallel$ and $\perp$ denote sternward resistance and lateral radiation force towards starboard (towards the right), respectively. 

The transient radiation forces may thus be evaluated by inserting $\zeta_t(\br,t)$ from \eqref{eq:tranW} into \eqref{R}, giving
\bs
\begin{align}\label{Rt}
  \begin{array}{c}R_{\parallel,t}(t)\\R_{\perp,t}(t)\end{array}&=-\frac \rmi{8\pi^2\rho}\lim_{\epsilon\to 0}\int_{-\pi}^\pi \rmd \gamma I(\gamma,t);  \\
  I(\gamma,t)&=\int_0^\infty \rmd k  \frac{k^3
  |p_\text{ext}(\bk,t)|^2  
  F(\bk)  }{\omega_\text{div}(\bk)} \left( \begin{array}{c} \cos\gamma\\\sin\gamma\end{array}\right) 
\notag \\
  &\times \left(\frac{\rme^{-\rmi\omega_+t}}{\omega_+-\rmi \epsilon}-\frac{\rme^{-\rmi\omega_-t}}{\omega_--\rmi \epsilon}\right)  .\label{I}
\end{align}
\es
Expressing radiation forces on the form \eqref{Rt} is useful for analytical purposes. For numerical purposes we use \eqref{R} more directly using a fast Fourier transform (FFT) method.

The static part of the wave resistance is obtained by inserting $\zeta_s$ into \eqref{R}. We refer to \cite{li16} for further details on the evaluation of the static part of the wave resistance.

\begin{figure}[htb]
  \includegraphics[width=.9\columnwidth]{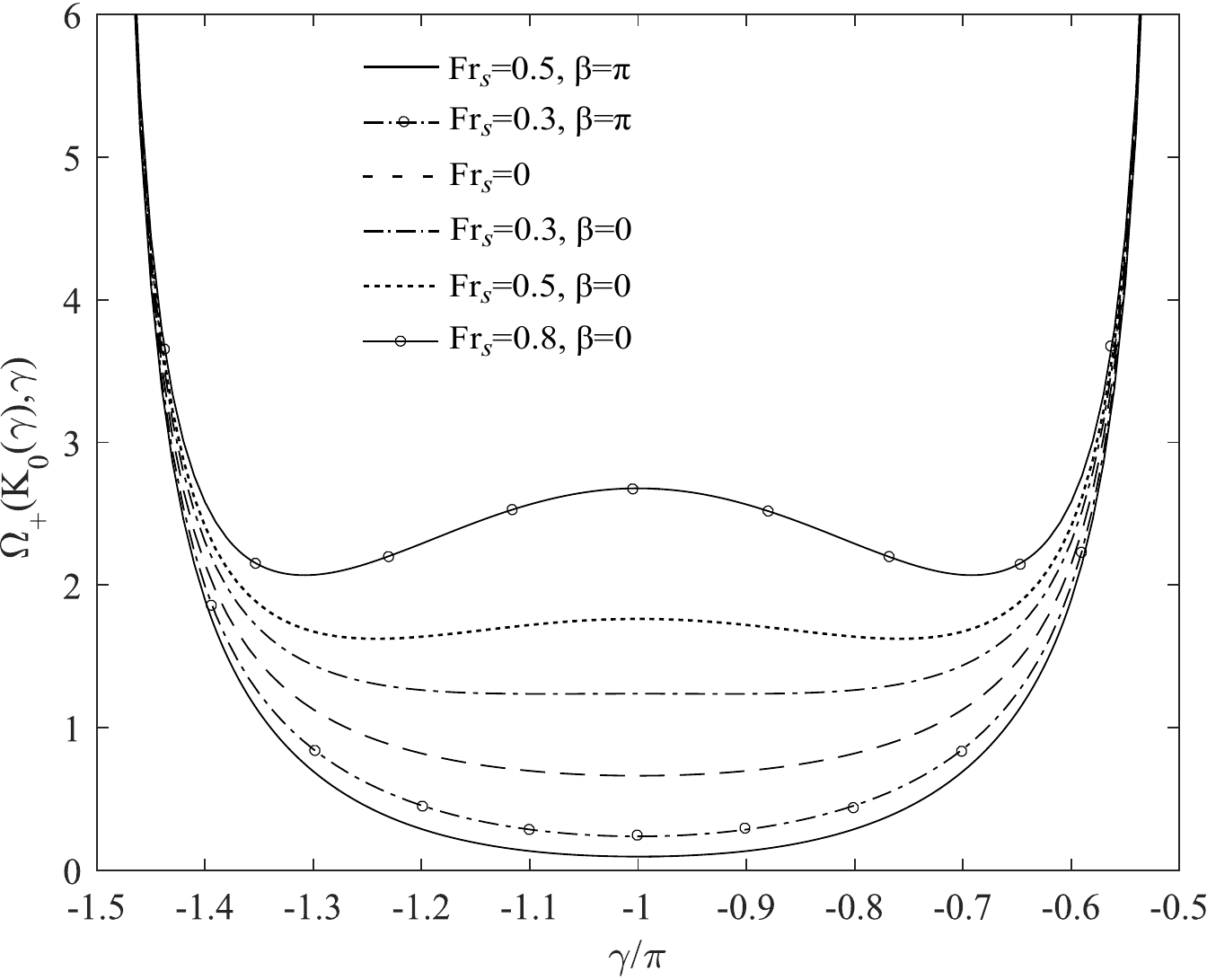}
  \caption{The intrinsic wave frequency $\tilde{\omega}_+(k_0(\gamma),\gamma)$, in units of $\sqrt{L/g}$, which solves the dispersion relation in direction $\gamma$ under different conditions of a linear shear profile. }
  \label{fig:omega}
\end{figure}

\subsubsection{Wave resistance oscillations}\label{sec:Rosc}

The transient behaviour of the wave resistance after the ship is set in motion, is to oscillate around its ultimate static value, at a frequency which varies greatly with direction or motion as well as shear strength. We will now explain what decides the oscillation frequency.  

The integral \eqref{I} is given solely by the contribution from the poles (infinitesimally close to) where $\omega_\pm(\bk)$ are zero. Since $\omega_+(\bk)$ and $\omega_-(\bk)$ are related through relation \eqref{omrel}, and we are free to replace $\bk\leftrightarrow-\bk$ under the integral sign, considering the zeros of the positive frequency $\omega_+(\bk)$ is sufficient. Taking the $k$ integral first as written out in \eqref{Rt}, the pole picks out a value $k_0(\gamma)$ so that
\be
  \omega_+(k_0(\gamma),\gamma) = 0.
\ee
Thus the particular frequency is picked out which satisfies the dispersion relation, which is to say that only waves which are able to propagate towards infinity along direction $\gamma$ may contribute to the wave resistance.

When $t$ grows large (while keeping $r$ constant), the exponential factor $\exp [-\rmi\omega_+(k_0(\gamma),\gamma)t]$ in the integrand of \eqref{I}, and is therefore dominated by the contribution from the value of $\gamma$ where the phase is stationary, that is, the value of $\gamma$ where
\be
  \partial_\gamma\omega_+(k_0(\gamma),\gamma) = 0.
\ee
Some time after $t=0$, the transient contribution to the wave resistance will therefore oscillate in time with the frequency of a stationary point, a maximum or minimum of $\omega_+(k_0(\gamma),\gamma)$ with respect to $\gamma$.

Let intrinsic frequencies be denoted with a tilde,
\be
  \tilde\omega = \omega-\kU_0 .
\ee
For the case of a linear shear current, we plot 
  $\tilde\omega_+(k_0(\gamma),\gamma)$ in units of $\sqrt{g/L}$  
as a function of $\gamma$ in Fig.~\ref{fig:omega}; 
$L$ is a characteristic length of the wave disturbance to be specified in particular examples below. 
We see that in all cases there is a stationary point at $\gamma=-\pi$. In the most shear--assisted direction ($\beta=0$), this frequency is enhanced compared to no shear, giving a faster oscillation of the wave resistance, whereas the opposite is the case in the maximally shear inhibited direction ($\beta=\pi$), where the oscillation can become very slow. For shear--assisted motion there are also two other stationary phase points at angles either side of $\gamma=\pi$, as is evident in Fig.~\ref{fig:omega}. Notably, the presence of shear which inhibits motion can dramatically decrease the oscillation frequency compared to still water, even at moderate shear.

\section{Numerical results}
\label{sec:num}

In this section we present numerical calculations of transient wave resistance on different model ships. While retaining generality by not specialising to particular real hull shapes, we have emphasised realism: a reasonably realistic model is used for the shape of the ship hull, and calculations are performed for a real velocity profile measured in the Columbia River estuary, where there is high traffic of vessels of many types. Parameters for vessel length and beam are taken from real ships known to travel in these waters.

The choice of the Columbia River delta for our data is primarily due to the excellent shear profile data available \cite{kilcher10}, although the location is also particularly apt for studies of ship wave effects. Thousands of ships ranging from carrier ships of more than $1000$ ft to small boats, are piloted up and down the Columbia river each year, in waters which are considered particularly trecherous, sometimes referred to as the Graveyard of the Pacific. 

Following \cite{li16,benzaquen14} we plot wave resistance relative to the constant
\[
 R_0 =  \frac{p_0^2}{2\pi^3 \rho g}.
\]

\subsection{Linear velocity profile}\label{sec:numlin}

In order to better highlight the underlying physics of the effect of shear on wave resistance, we begin by considering the simplest shear flow, which varies linearly as a function of depth, $U(z)=U_0+Sz$. Realistic shear profiles are considered in Section \ref{sec:genprof}. 
\begin{figure*}[tb]
  \begin{center}
    \includegraphics[width=.8\textwidth]{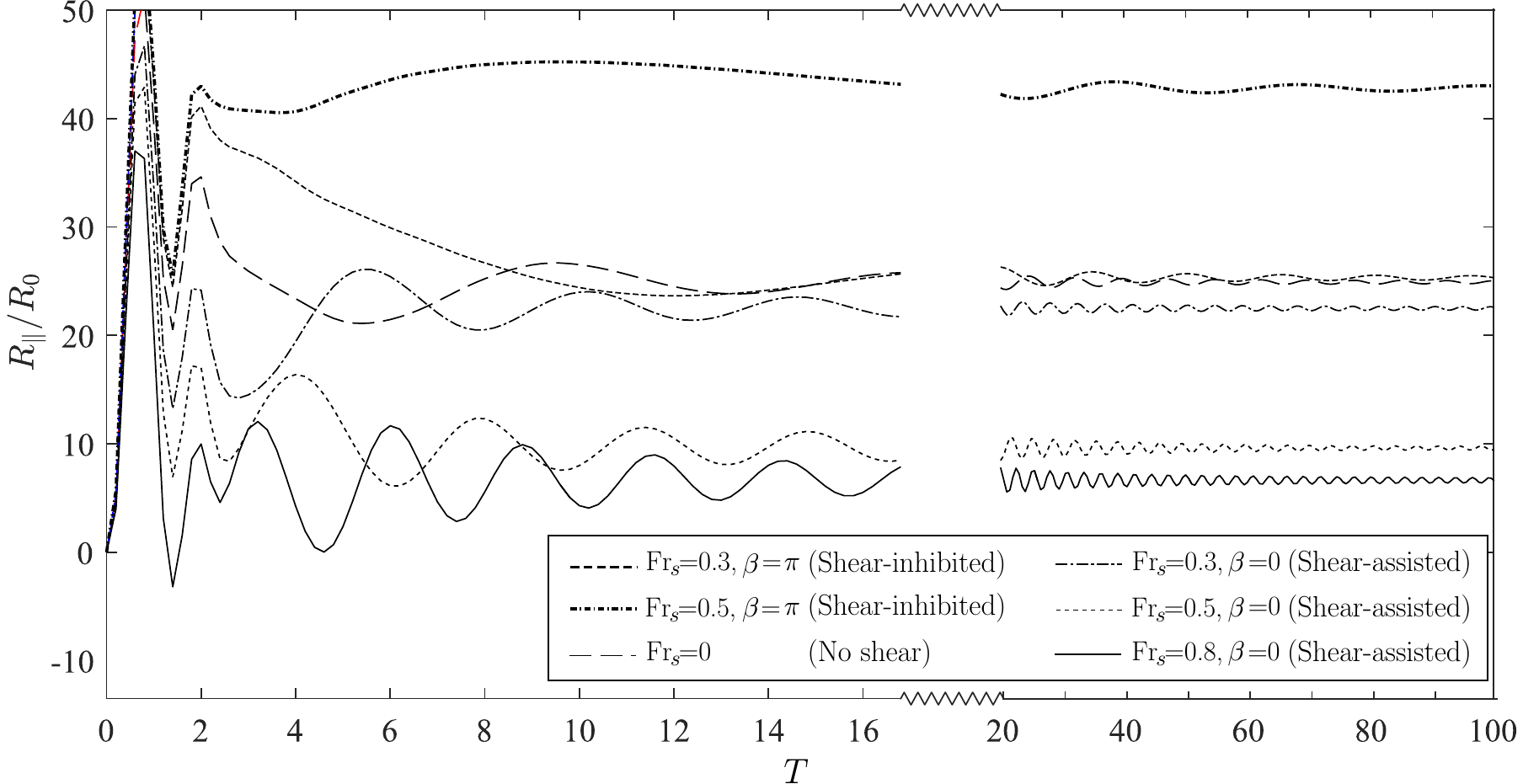} 
  \end{center}
  \caption{Transient wave resistance on a ship set suddenly in motion as a function of nondimensional time $T=t\sqrt{L/g}$, for different cases where a linearly depth-dependent shear current is present in deep water. 
  The ``ship'' is modelled as an ellipsoidal, super-Gaussian surface pressure of  aspect ratio $6$ and $L=1$ (arbitrary units),   moving with $\Fr =0.3$. Note that the abcissa is scaled differently for $T>20$. Note furthermore that $R_\parallel$ scales linearly with $L$, which is arbitrary in this scale--free system, hence so is the scaling of the ordinate axis.}
  \label{fig:resistance}
\end{figure*}
\begin{figure}[tb]
  \includegraphics[width=\columnwidth]{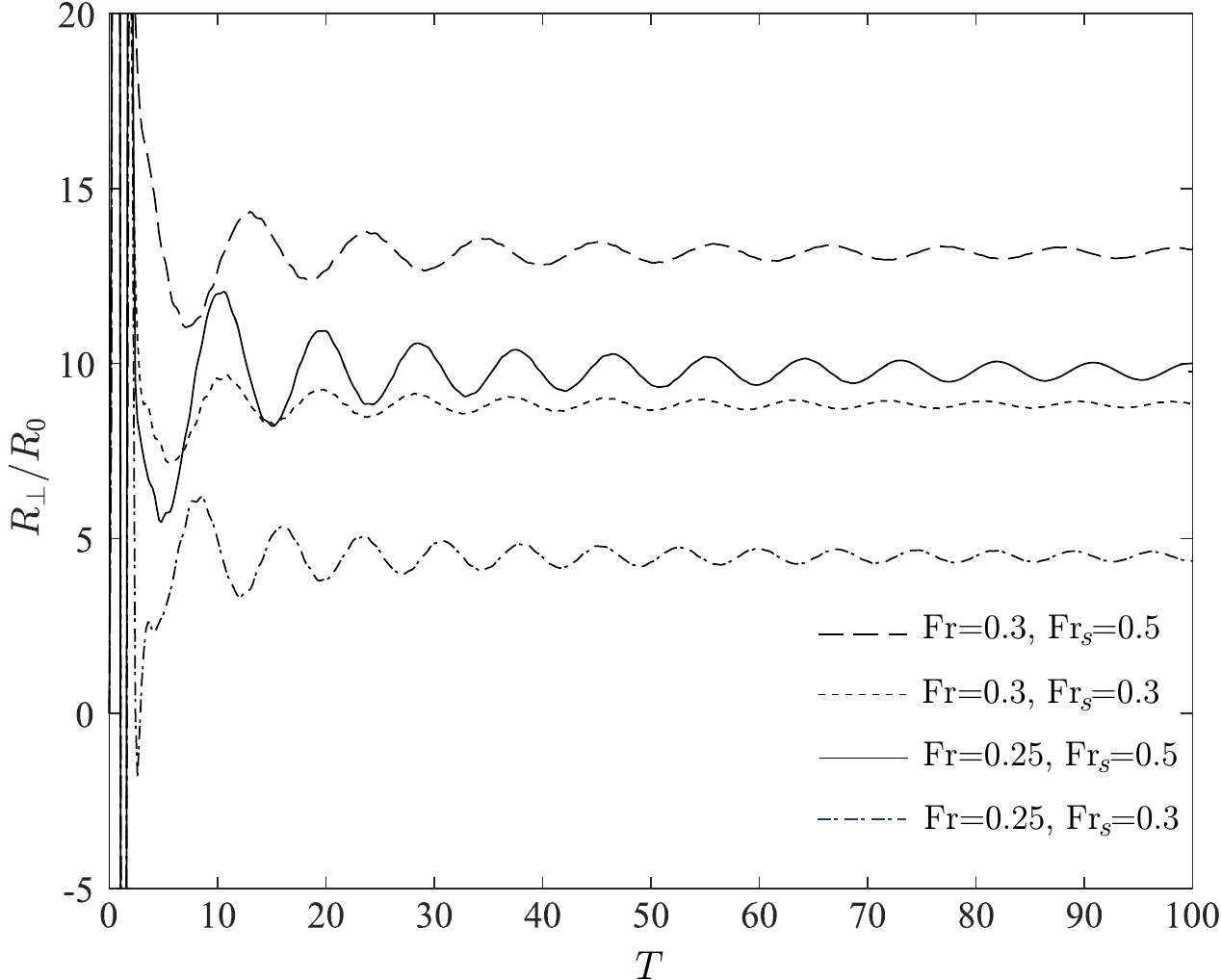}
  \caption{Same as Fig.~\ref{fig:resistance}, but for the transient lateral wave radiation force  $R_\perp$, for motion normal to a the shear current in a reference frame following the water surface; $\beta=\pi/2$ (see Fig.~\ref{fig:geom}). The scaling of the ordinate is arbitrary (see Fig.~\ref{fig:resistance}).}
  \label{fig:lateral}
\end{figure}

\subsubsection{Suddenly starting ship}

For the simplest, linearly varying velocity profile considered in section \ref{sec:linear} we calculate the transient wave resistance for a ship modelled as in equation \eqref{p}, 
whose velocity goes suddenly from zero to a constant value $V$. While idealised, this models a starting ship without the need for further parameterisation of the acceleration phase. An example of what the transient wave resistance  looks like is shown in figure \ref{fig:resistance}. The model ship is elliptical with aspect ratio $6$ and length $L=1$ (arbitrary units since the problem is intrinsically scale--free), and calculation is performed for $\Fr =0.3$ and shear strengths varying from $\Fr _s = 0$ to $0.8$.

The oscillation frequencies of the transient wave resistance are found to agree well with the stationary phase values of $\omega_+(k_+(\gamma),\gamma)$ in figure \ref{fig:omega} as expected. 

The transient wave resistance is seen to go through a sharp peak shortly after the ship is set in motion, and then relax in an underdamped manner towards its steady--motion value. For shear--assisted motion  ($\beta=0$), the initial peak can be much higher than its static value, whereas this effect is weaker in the case without shear ($\Fr _s=0$) and for shear--inhibited ship motion. Letting the shear vary from strongly inhibiting (high $\Fr _s$, $\beta=\pi$) via no shear to fairly strongly assisting, we see that the transient oscillations increase both in amplitude and frequency, whereas the static wave resistance decreases. An interesting observation is that for very strongly motion-assisting current ($\Fr _s=0.8$ in this case), the total wave resistance can actually be negative during some time intervals, since oscillation amplitudes are large and the static wave resistance correspondingly small. 

Both the difference in oscillation frequency and the magnitude of the steady motion wave resistance can be understood by considering the relative values of phase velocity and group velocity in different directions of wave propagation. A detailed discussion of this may be found in Ref.~\cite{ellingsen14b}. For a linear shear current, where the dispersion relation \eqref{linrel} is known analytically, one finds that in a reference frame following the free surface, the group velocity is quite similar in all directions of motion, whereas phase velocity can differ greatly. In shear--inhibited directions dispersion is weakened and an emitted wave group will retain its initial shape and width to a greater extent than in quiescent water. The opposite is the case for shear--assisted wave propagation; here the phase velocity can far exceed the group velocity, so wave groups quickly spread and have a rapidly changing, volatile appearence. 

When the ship suddenly starts, an initial ring wave is emitted, as seen in Fig.~\ref{fig:patterns}. Wave resistance will continue to oscillate for as long as this ring wave remains in the ship's near--zone. The fact that group velocity is fairly isotropic means that it takes approximately the same time for the ring wave to disappear from sight, matching the observation that the oscillations in Fig.~\ref{fig:resistance} die off at a similar rate in all cases. The frequency of oscillation, however, depends on the phase speed of the transient waves \emph{within} the ring wave group, and the higher phase velocity for shear--assisted propagation means faster oscillations, as also observed in Fig.~\ref{fig:resistance}, and explained in connection with Fig.~\ref{fig:omega}. 

Finally, we found in Ref.~\cite{li16} that the effect on shear on wave resistance is, in a rough sense, to effectively change the Froude number to a value based on the ship velocity relative to some depth--average current speed rather than its surface value. 
The Froude number is effectively lowered in shear--assisted motion, and increased in shear--inhibited motion. 
A detailed discussion is found in section \ref{sec:realsudd} where we compare a real velocity profile to a linear approximation in this respect. Since the general trend is that wave resistance increases with increasing $\Fr $ for $\Fr \sim0.3$, this explains why the resistance in steady motion is typically decreased for shear--assisted motion and increased for shear--inhibited motion. However, this does not always hold true, due to interference effects between waves from bow and stern. 

We go on to calculate the transient lateral radiation force for the same ship, shown in Fig.~\ref{fig:lateral}. 
The Froude numbers $0.25$ and $0.3$ are chosen as realistic examples. The ship motion is now across the shear current, $\beta=\pi/2$ as defined in figure \ref{fig:geom}. For an aspect ratio of $6$ the lateral force is roughly half the magnitude of the sternward force. We find the relative magnitude of lateral to sternward force to vary strongly with Froude number and aspect ratio, as indicated for the former case by the large effect of lowering $\Fr $ from $0.30$ to $0.25$.

\subsection{General, realistic velocity profiles}\label{sec:genprof}

\begin{figure}[tb]
  \includegraphics[width= .9\columnwidth]{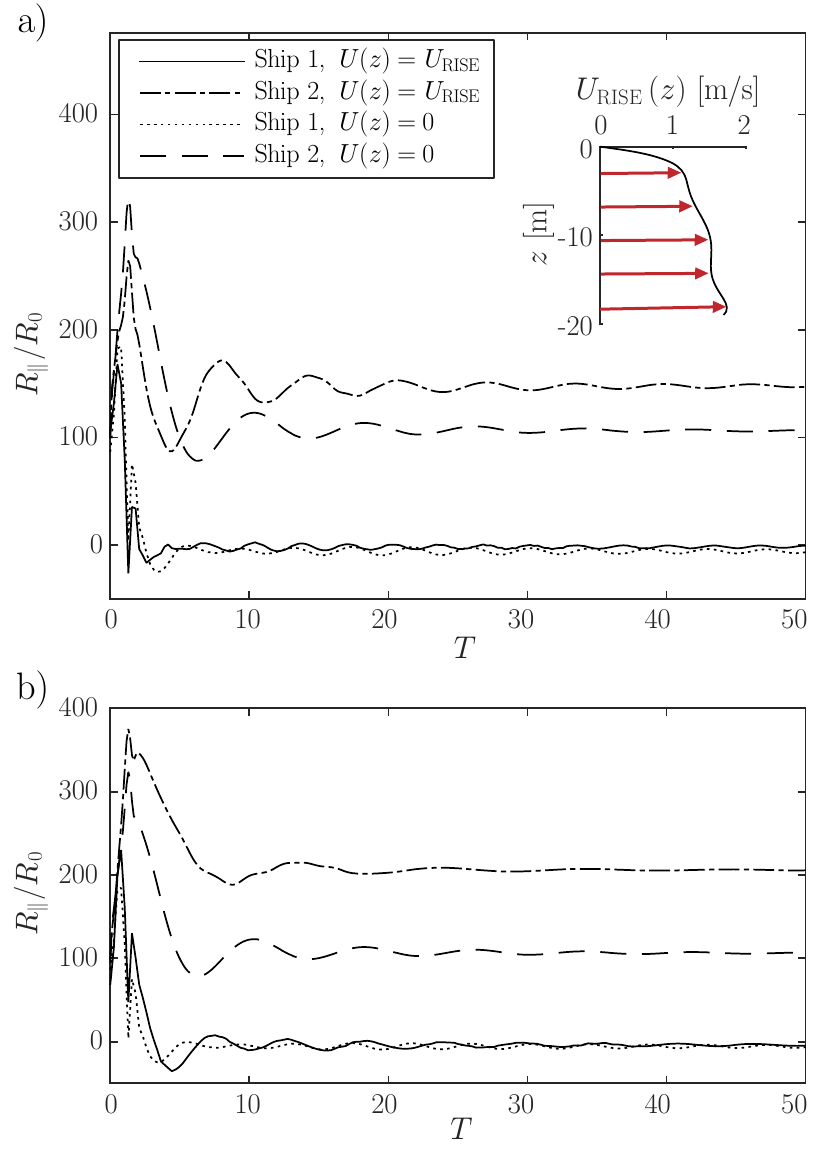}
  \caption{Transient wave resistance for motion in the shear assisted (a) and inhibited (b) directions atop the measured current in the Columbia River delta, as a function of nondimensional time $T=t\sqrt{g/L}$. Three ships are modelled with equation \eqref{p} with dimensions as given in Table \ref{tbl:ships}. 
  The wave resistance in quiescent waters is shown for comparison. Inset to (a): measured Columbia River velocity profile $U_{\mathrm{RISE}}(z)$ \cite{kilcher10} approximated with a $6$th order polynomial, in a reference frame moving with the surface current. The legend applies to both a) and b).}
  \label{fig:R0}
\end{figure}

\begin{figure}[tb]
  \includegraphics[width=.9\columnwidth]{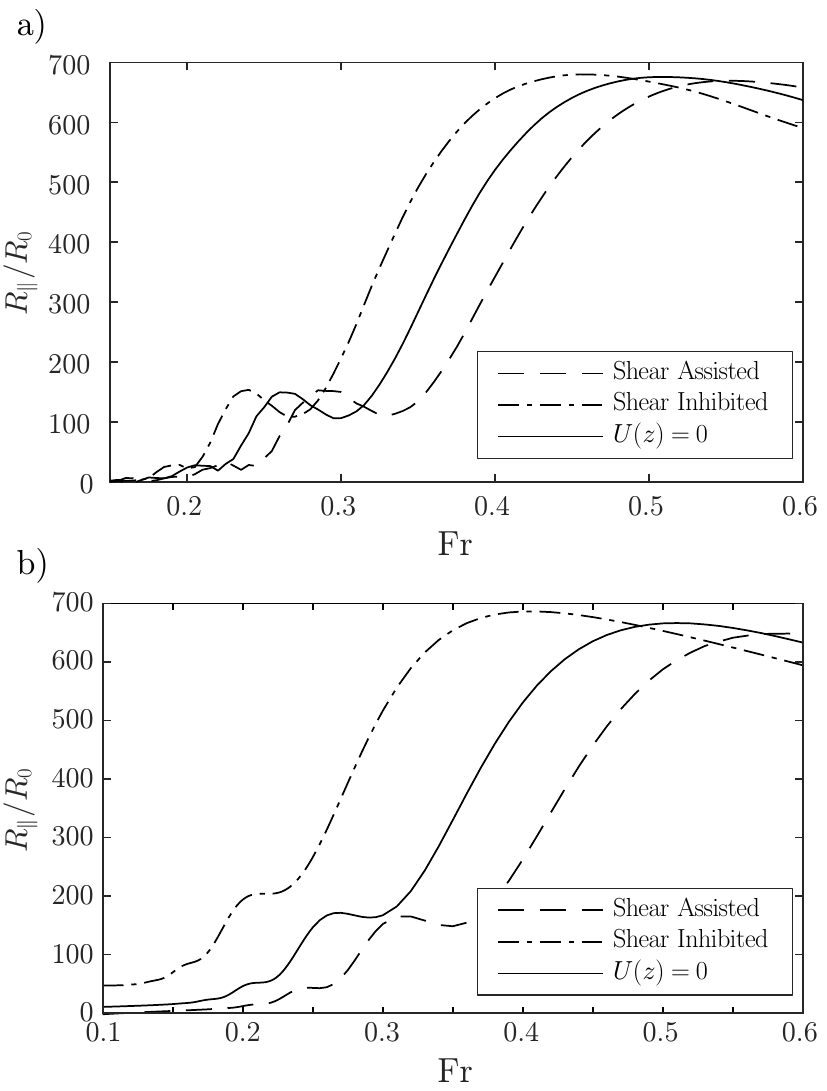}
  \caption{Wave resistance force in steady motion for Ship 2 (tugboat) as a function of Froude number for the maximally shear assisted ($\beta=0$) and inhibited ($\beta=\pi$) directions of motion. a) The Columbia River velocity profile, b) Linearly varying profile $U(z)=U_0+Sz$ with $\Fr_s\equiv U_0S/g=0.4$.}
  \label{fig:RofF}
\end{figure}

We now compute the transient wave resistance using a real, measured velocity profile. The shear current is that measured by the RISE project, a tidal current in the mouth of the Columbia River \cite{kilcher10}\footnote{Since measurements begin at $2$m depth, we presume this point to be at the surface, thus offsetting all data by $2$m. This should be a conservative procedure since shear strength increases closer to the surface.}. Buoyant fresh water from the river creates a strong surface jet as it enters the salt water of the Pacific Ocean. We approximate the measured data with a 6th order polynomial which is then subjected to the piecewise-linear procedure to calculate the dispersion relation numerically, as described in section \ref{sec:basic} and detailed in \ref{appx-2}. 
The current profile $U_{\mathrm{RISE}}(z)$ in a reference frame where the surface current is zero is shown in the inset of figure \ref{fig:R0}a. We model various ships using Eq.~\eqref{p} with dimensions $L$ (length) and $b$ (beam) representative of typical vessels traveling at the Columbia River mouth, tabulated in Table \ref{tbl:ships}.

\begin{table}[tb]
\begin{center}
\begin{tabular}{cllllll}
\hline
ID&Ship Type&Length & Beam  & Speed & Aspect \\ 
&& $L$ [m] &  $b$ [m]  & [Knots] & ratio\\ \hline
1& Bulk carrier & 170 & 28 & 11.9 & 6.07 \\ 
2& Tugboat  & 32 & 10.4 & 10.3 & 3.08 \\ 
3& Fishing boat & 19 & 6 & 8.0 & 3.17 \\ \hline
\end{tabular}
\caption{Parameters of the modeled ships, chosen as representative dimensions from boat traffic on the Columbia River. Real-time data on vessels in these waters is available at http://www.columbiariverbarpilots.com. Froude numbers for ships $1,2,3$ are $0.15,0.3$ and $0.3$, respectively.}
\label{tbl:ships}
\end{center}
\end{table}

\subsubsection{Suddenly starting ship}\label{sec:realsudd}

Results for transient sternward wave resistance for a ship starting suddenly in maximally shear-assisted and shear-inhibited directions of motion (corresponding to upstream and downstream motion in the Columbia delta, respectively) are shown in figure \ref{fig:R0}. 
Two ships are modelled, 
a bulk carrier ship, and a smaller vessel typical of a tugboat; Ships 1 and 2 in Table \ref{tbl:ships}, respectively. 
The wave resistance in quiescent waters is shown for comparison.
 
The behaviour 
of the smaller ship ('Ship 2')
is similar to that observed for the simple linear shear current, with wave resistance exhibiting a sharp peak shortly after the ship is set in motion, whereupon it relaxes in an underdamped way to the steady motion value with a frequency which is higher for shear assisted than for shear inhibited motion. Fluctuations are stronger for shear assisted (upstream) motion as was also noted in Fig.~\ref{fig:resistance}, and amount to transient variations in the order of $10\%$ of the static value in this case. The wave resistance of the larger vessel (`Ship 1') approaches an insignificantly small value at large times, attributed to the lower Froude number ($0.15$) for this modelled vessel. 

The most interesting observation made in Fig.~\ref{fig:R0} might concern the steady motion value of wave resistance. Untypically, wave resistance is increased compared to quiescent waters both for shear--assisted and shear--inhibited ship motion. This appears to run counter to lessons learned from a previous, much simpler and less realistic model study \cite{li16}, where shear--assisted motion was always found to decrease wave resistance in this Froude number range. The reason is that our present, more realistic ship model \eqref{p} has a sharper bow and stern than the circular ``ship'' considered in \cite{li16}, leading to interference effects between bow and stern waves such as are found for real ships. Indeed these interferences must be taken into account when choosing optimal operational speed in ship design \cite{schneekluth87}. We plot the Froude number dependence of the steady motion wave resistance for different $\Fr$ for the Colubia current profile in Fig.~\ref{fig:RofF}a for the tugboat (Ship 2). The plot clearly demonstrates that wave resistance in steady motion depends very strongly on direction and Froude number. $\Fr =0.3$, the speed of Ship 2 in Fig.~\ref{fig:R0}, is a special case where shear increases wave resistance in both directions. Increasing the velocity a little to $\Fr =0.33$, a very different conclusion is reached: here, shear-inhibited wave resistance (ship travelling downstream) is more than a factor $3$ greater than in the opposite direction. 

\begin{figure}[htb]
  \includegraphics[width= .9\columnwidth]{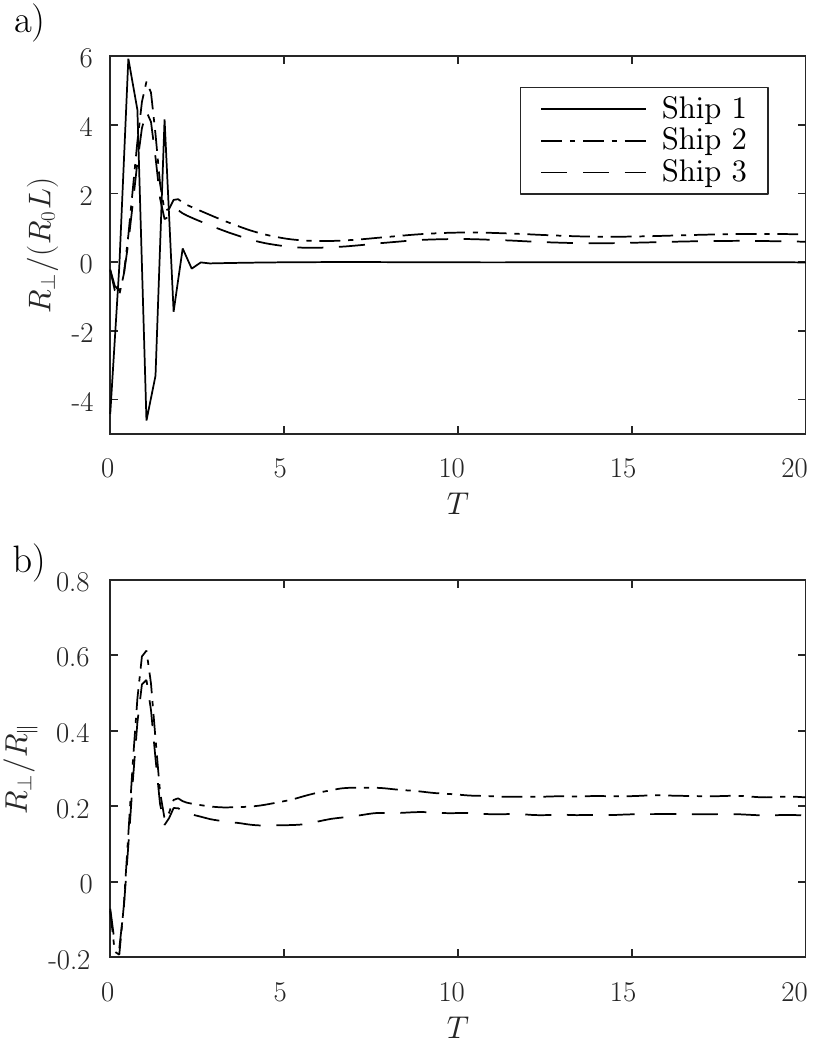}
  \caption{
    a) Transient lateral radiation force per unit ship length $R_\perp/(R_0L)$ for motion normal to the measured current in the Columbia River delta, in a reference system where the free surface is at rest ($\beta=\pi/2$), as a function of nondimensional time $T=t\sqrt{g/L}$. Three ships are modeled using \eqref{p} with dimensions $L$ and $b$ and Froude number $Fr$ as indicated. b) Transient lateral radiation force relative to transient wave resistance for the two smaller modeled ships. The legend applies to both a) and b).
  }
  \label{fig:R90}
\end{figure}

\begin{figure*}[htb]
  \includegraphics[width= 2.0\columnwidth]{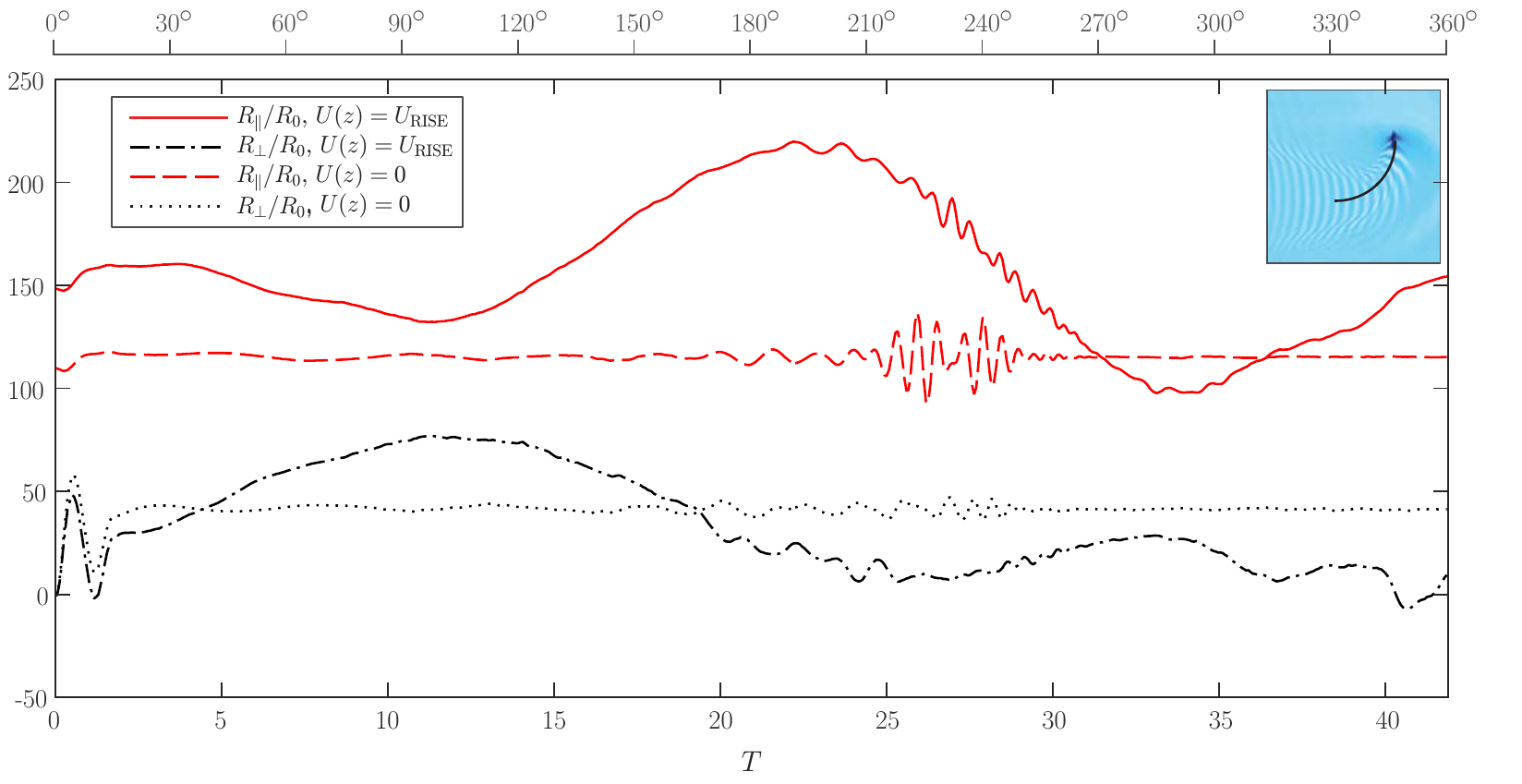}
  \caption{
    Transient wave resistance as a function of nondimensional time $T=t\sqrt{g/L}$ for a ship beginning a circular manoeuvring motion atop the measured shear current in the Columbia River delta. The ship dimensions are typical of a tugboat operating in these waters, `Ship 2' in Table \ref{tbl:ships}, initiating a turn of radius $4L$ at $T=0$, from having traveled in a straight path in the shear--assisted direction (upstream). The path is circular thereafter, as seen in a reference system where the water surface is at rest. The angle the ship has turned is shown above the figure. The situation at $90^\circ$ is shown in the inset for illustration. Also shown is the same manoeuvre in quiescent waters ($U(z) = 0$). 
  }
  \label{fig:turn90}
\end{figure*}

Corresponding results for the lateral radiation force for motion across the shear current ($\beta=\pi/2$, measured in a reference frame in which the water surface is at rest) is shown in figure \ref{fig:R90} 
for the three different ships in Table \ref{tbl:ships}. In order to make the values comparable, we divide the force by the length of the ship. The lateral radiation force shows similar oscillations for short times as the sternward resistance in Fig.~\ref{fig:R0}a, with the exception of the large carrier ship (Ship 1) which displays far stronger transient oscillations initially. Indeed, while the sternward resistance force is likely to be negligible for Ship 1, this needs not be the case for the early transient shortly after start. 

In Fig.~\ref{fig:R90}b we show the lateral radiation force relative to the sternward resistance force for cross--current motion, for Ships 2 and 3. The relative strength has very weak oscillations, but a highly conspicuous trait is how the relative strength of the transient force is more than twice as strong just after appearance of the ``ship'' compared to its asymptotic value, about $50-60\%$ percent of the transient sternward force at the time of the initial peak that is present in both force components. Again this indicates that the transient behavior of the lateral force could well have a bearing on seakeeping performance during manoeuvering, when transient waves will be emitted by the ship. 
We note furthermore that when stationary conditions have been reached, the radiation force is approximately $20\%$ of the sternward component. This is a significant laterally directed force which must be compensated by steering (it is not to be confused, of course, with the lateral drag force which will also be present due to the shear flow between surface level and the ship's draught, a separate question not studied here. With no shear there is neither a net lateral drag nor radiation force when $\beta=\pi/2$.)

The simplicity of working with the linearly dependent velocity profile as a model for a real current makes it tempting in practice to eschew the need to calculate $\omega(\bk)$ for a general shear flow, and instead approximate the real profile by a linear one with a representative constant shear. However, if we were to approximate the Columbia profile by a linear profile with a shear approximately that at the water surface --- giving $\Fr _s\sim 0.4$ for our parameters --- one could make a very great error in calculating the steady-motion wave resistance. In Fig.~\ref{fig:RofF}b we plot the steady--motion wave resistance as a function of $\Fr$ using this model. It is clear that while the trend and general behaviour is similar, the rapid variation of $R_\parallel$ with $\Fr$ for $0.2\lesssim\Fr\lesssim 0.4$ means the error can be several hundred percent. Clearly a better job can be made with a better choice of $\Fr_s$, yet choosing a sufficiently good value in practice (if such exists) will require the use of knowledge of the full velocity profile and moreover be specific to each vessel. In our opinion this may not be any simpler nor numerically cheaper than a full calculation such as we have performed, and for which an effective calculation tool is already now developed. 

We note, however, the possibility that a two--layer model might be a compromise which is the best of both worlds. In such a model a surface layer is given one constant shear value, and deeper waters another. It is well suited for modelling a surface shear layer due to wind or tides for many practical purposes. Such a model is analytically tractable while containing the key parameter of the vertical extent of the surface shear layer, whose relation to the ship length is a determining parameter. Analysis of such a model in the context studied here is beyond our present scope; the dispersion relation that can be used directly in the formalism of Section \ref{sec:basic} may however be found in Ref.~\cite{smeltzer17b}.

\subsubsection{Turning ship}

Analysis of a suddenly moving ship yields insight into transient wave resistance forces
due to sudden changes in velocity along a straight course. 
It is of interest to consider another 
example of a ship manoeuvre; a turning motion. Figure \ref{fig:turn90} shows the wave resistance for a ship initially traveling along a straight path upstream in the Columbia River delta (shear--assisted direction), which begins a circular turning manoeuvre of radius $4L$ at $T = 0$. The forward velocity $\Fr=0.3$ remains unchanged through the manoeuvre. 
We consider as example a typical tugboat operating in these waters, Ship 2 in Table \ref{tbl:ships}. 
In order to given an impresson of all different directions of motion, we let the ship do a full $360^\circ$ turn; a snapshot at $90^\circ$ is shown in the inset. The same ship manoeuvre in quiescent waters is shown for comparison. 

All graphs display certain oscillations at different times during the manoeuvre, due to the sudden change in lateral acceleration after $T=0$, and later because the ship encounters its own previously emitted waves.

In quiescent water the lateral radiation force fluctuates around a constant value of, in this case, approximately $40R_0$ due to the now asymmetric wave field; another way of seeing it is that the turning ship must accelerate water towards the centre of the arc, resulting in an outwardly directed lateral added mass force. The sternward force without shear also fluctuates around a constant as it should. 
A different behaviour is observed for both force components, however, when the measured Columbia River shear current is present. Both resistance and lateral force vary greatly throughout, both peaking at around twice their quiescent value, and the lateral force at times dropping to zero and even small negative values. For a ship to follow such a path with precision will thus require considerably greater skill than in quiescent water, having to account for the changing lateral and sternward forces. The lateral force can also reach more than $50\%$ of the resistance force for a part of the circle with our parameters, typical of boat traffic in the area, by no means a small force in a manoeuvring context.

\section{Conclusions}

We have studied the wave radiation forces, including wave--making resistance, for different model ships in a real, measured current in the Columbia River delta. We calculate transient wave resistance on a ``ship'' modeled as a traveling pressure distribution in the form of an elliptic super--Gaussian.
Choosing values of length/beam typical of smaller vessels (tugboats, fishing boats) we find that wave resistance can vary drastically depending on direction of motion, upstream or downstream, showing a strong dependence on Froude number. For typical Froude numbers --- $\Fr\sim 0.2$ to $0.4$ ---  we find that wave resistance can differ by more than a factor $3$ between upstream and downstream motion. Appropriate choice of vessel velocity can thus make a large difference to resistance in strongly sheared waters. 

When there is an oblique angle between the ship's line of motion and the shear current, the emitted ship wave pattern will be asymmetric, with more waves propagating to one side than the other. The total wave radiation (or wave--making) force then also has a lateral component. For our example model ships representative of tugboats or fishing boats, the lateral force was found to be approximately $20\%$ of the sternward resistance force for a ship in steady motion. 

We also study the transient behaviour of wave radiation forces acting on ships which change their velocity. As a simple example we consider ships that are set suddenly in motion. Both components of the wave radiation force undergo an initial peak as an initial ring wave is created, whereupon they oscillate in an underdamped manner towards their steady--motion values. For motion across the shear current the lateral force is found to have a stronger initial peak, and the lateral force momentarily reaches more than 50\% of the value of the sternward force just after motion commences. 

The general trend for typical small--ship operational Froude numbers is that compared to quiescent water, wave resistance decreases for upstream (shear--assisted) ship motion, and increases for downstream (shear--inhibited) motion, although interference effects between bow waves and stern waves can alter this for certain Froude numbers.

We also considered a circular manoeuvring motion atop the Columbia River current seen from a reference system following the water surface,  for a small ship (tugboat). Unlike on quiescent water were both resistance and lateral force are constant through the motion (modulo small oscillations due to encountering the ship's own waves), these vary greatly through the circular path on the Columbia River mouth. Variations of amplitude of approximately $100\%$ of the quiescent values of the forces are found. For a ship to follow such a path with precision will thus require considerably greater skill. The lateral force can also reach more than $50\%$ of the resistance force for a part of the circle with our parameters, typical of boat traffic in the area.

The second main achievement reported in this manuscript is the development of a theory that allows calculation of waves from a general, time-dependent applied surface pressure acting on the free surface of a horizontally directed shear current which may vary arbitrarily with depth in both direction and magnitude. We present a framework which provides the means to effectively calculate ship waves and wave resistance without undue difficulty. The theory is based on deriving the response of a water surface satisfying an arbitrary dispersion relation, to an impulsive applied pressure. The wave pattern is then calculated as the integral of emitted waves at all previous times. It is necessary to devise a scheme to obtain the dispersion relation numerically; in this paper we used the piecewise--linear approximation \cite{smeltzer17}, but several other options are available. 

\subsection*{Acknowledgements}

S{\AA}E is funded by the Norwegian Research Council (FRINATEK), project number 249740. We are grateful to Peter Maxwell for improvements to the PLA numerical code.

\appendix
\section{Derivation details}

\subsection{The implicit dispersion relation} \label{appx-1}

We here derive in detail the implicit dispersion relations \eqref{dispR} and \eqref{eq:dwF}. We first make the ansatz that for a progressive wave of oscillating frequency $ \omega $ and wave vector $\bk$, $w$ and $p$ are of the following form,
\be \label{eq:defwp}
[w(\br,z,t), p(\br,z,t)] = [\tilde{w}(\bk,z),\tilde{p}(\bk,z)]\rme^{-\rmi \omega t+\rmi\bk\cdot\br}.
\ee
Eliminating $\tilde{p}$ after inserting \eqref{eq:defwp} into \eqref{eq:p} and \eqref{eq:dp} yields the Rayleigh equation
\be
(\kU-\omega)(\p^2_z-k^2)\tilde{w} = \kU'' w,
\ee

We define  $H_w(\bk,z) =\sinh k(z+h)/\sinh kh$, and notice that since $(\partial_z^2-k^2)H_w=0$, 
\begin{align} \label{eq:gre}
&\zint \left[ H_w(\bk,z) (\p^2_z-k^2)\tilde{w} +\tilde{w}(\p^2_z-k^2) H_w(\bk,z) \right]  \notag\\
& = H_w(\bk,0) \tilde{w}'(\bk,0) - w(\bk,0) H'_w(\bk,0)\notag\\
& =\zint \dfrac{\kU''  \tilde{w} H_w(\bk,z) }{\kU-\omega},
\end{align}
where the seabed condition $ \tilde{w}(\bk,-h) = 0 $ was applied. 
The homogeneous boundary condition for $w$ at the free surface is found as
\be \label{eq:bcsf}
 (\kU_0-\omega)^2 \tilde{w}_0\prime-[\kU_0\prime(\kU_0-\omega) +gk^2 ]\tilde{w}_0 = 0,
\ee
where the subscript $0 $ denotes the values at $ z=0 $. 

Inserting the condition \eqref{eq:bcsf} into \eqref{eq:gre} then yields \eqref{dispR}. Moreover, \eqref{eq:bcsf} and the dispersion relation \eqref{dispR} further give \eqref{eq:dwF} since $ w' = \tilde{w}' $ at $ t = 0^+ $. 

\subsection{The piecewise linear approximation} \label{appx-2}
\begin{figure}[htb]
  \includegraphics[width=\columnwidth]{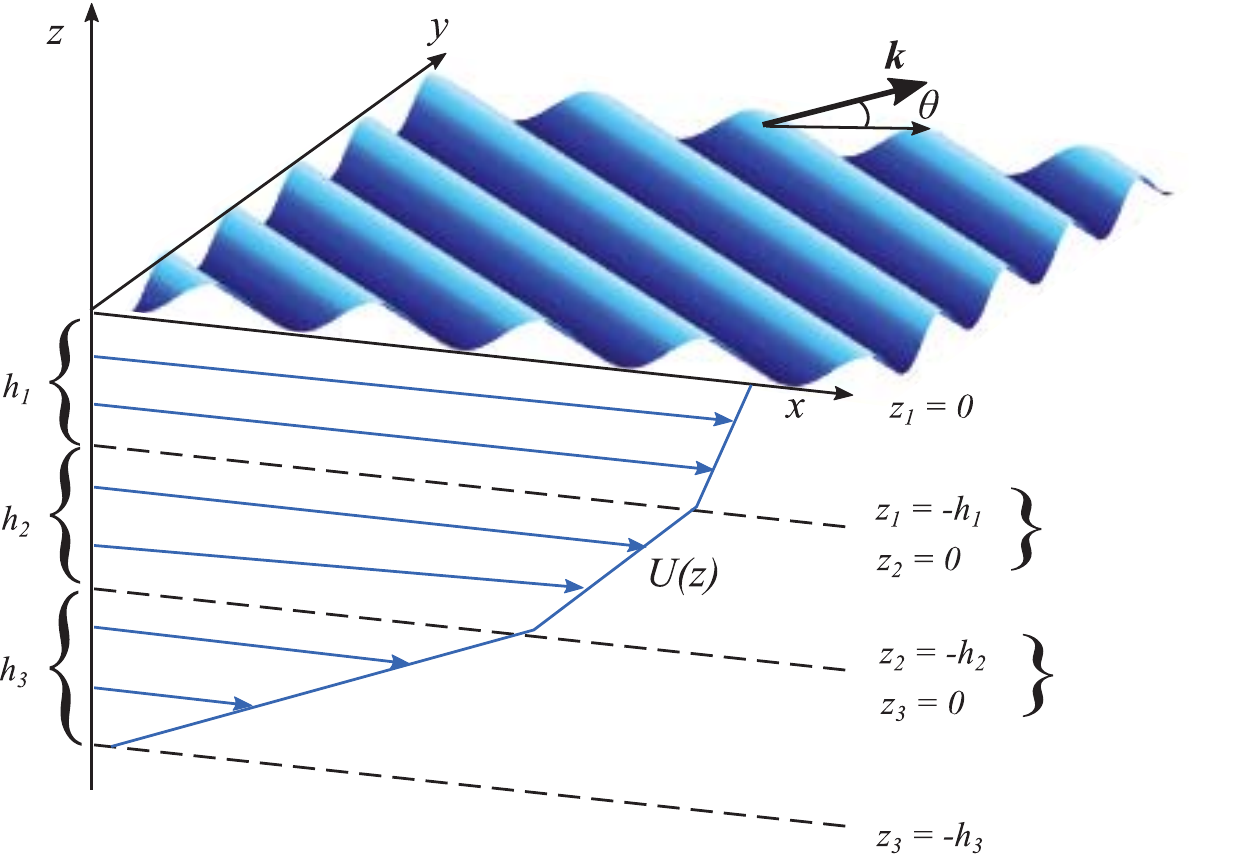}
  \caption{Piecewise linear approximation (PLA): The velocity profile approximated by a piecewise-linear function, dividing the fluid ito $N$ artificial layers.}
\label{fig:nlayer}
\end{figure}
Following \cite{smeltzer17}, the fluid is artificially divided into $N$ layers in the vertical direction each with thickness $h_j$ and constant vorticity $S_j$ as shown in Fig.~\ref{fig:nlayer}. A vertical coordinate $z_j$ is defined within each layer where $z_j = 0$ and $ -h_j $ at the top and bottom layer interfaces respectively. The approximate piecewise linear background velocity profile in each layer is 
\[
  U_j^{PL} = U_{j-1} + S_jz_j. 
\]

A spatially uniform velocity $V_0$ in the $y$ direction, corresponding to a translation of the frame of reference, can be added when needed, amounting only to the addition of a Doppler shift $k_y V_0$ to the wave frequencies $\omega(\bk)$ as calculated with the PLA. 

As presented in more detail in \cite{smeltzer17}, solutions $u_j,v_j,w_j,p_j$ to the linearized Euler equations can now be found within each layer $j=1,2,...,N$ modulo undetermined coefficients, and solutions are matched by requiring continuity of $w$ and $p$ (kinematic and dynamic boundary conditions, respectively) across the artificial layer boundaries, as well as free surface boundary conditions at $z=0$ and vanishing $w$ at $z=-h$ (or $z\to -\infty$). To wit one obtains
\begin{subequations}  \label{eq:nlayer}
\begin{align}
(\p_t+\rmi\kU_j)&(\p^2_{z_j}-k^2)w_j =0,~ 
                                       -h_{j}<z_j<0,\label{na} \\
p_1-\rho g\zeta & = \pext, ~\text{at}~z_1 =0, \label{eq:bc0}\\
w_1                 &  = (\p_t+\rmi\kU_1)\zeta,~~\text{at}~z_1 =0, \\
w_j(z_j=-h_j)  &  = w_{j+1}(z_{j+1}=0),\label{nb}\\
p_j(z_j=-h_j)  &  = p_{j+1}(z_{j+1}=0),\label{eq:bc4}\\
w_{_N}         &  =0, ~z_{_N} =-h_{_N}.  
\end{align}
\end{subequations}
where \eqref{na} holds for $1\leq j\leq N$, and \eqref{nb} and \eqref{eq:bc4} hold for $1\leq j\leq N-1$.

In particular, the vertical velocity perturbation and the dynamic pressure distribution are of the following forms, respectively
\begin{subequations} \label{eq:wjpj}
	\begin{align}
	w_j    &= A_j(\bk,t)\sinh{k(z_j+h_j)}\notag\\& + B_j(\bk,t)\cosh{k(z_j+h_j)},\\
	-kp_j/\rho &=(\p_t+\rmi \bkU_j){w'_j}-{\rmi k_x S}w_j.
	\end{align}  
\end{subequations}
Inserting \eqref{eq:wjpj} into \eqref{eq:nlayer} and eliminating the $B$ coefficients yields set of $N+1$ linear equations. The eigenvalues of $\omega(\bk)$ are found from requiring the determinant of the system matrix be zero, the criterion for nontrivial solutions of the homogeneous system to exist. This gives, in general $N+1$ eigenvalues, of which two are physical and an appropriate procedure must be employed to choose the correct values, as detailed and discussed in \cite{smeltzer17}. 
The procedure moreover automatically provides the coefficients $A_1$ and $B_1$ required in Eq.~\eqref{FPLA}.

\section*{References}

\end{document}